# Does Predictive Performance Translate to Causal Performance?

Evaluating tabular foundation models as nuisance learners for causal effect estimation

**Authors:** Andrew R Weckstein[1,2,3,*], Issa J Dahabreh[2,3,4,5], Marinka Zitnik[6,7], Sonia Hernández-Díaz[2,3], Sebastian Schneeweiss[1,2]

**Affiliations:**

1. Division of Pharmacoepidemiology and Pharmacoeconomics, Department of Medicine, Brigham and Women's Hospital and Harvard Medical School, Boston, MA, USA
2. Department of Epidemiology, Harvard T.H. Chan School of Public Health, Boston, MA, USA
3. Department of Biostatistics, Harvard T.H. Chan School of Public Health, Boston, MA, USA
4. CAUSALab, Department of Epidemiology, Harvard T.H. Chan School of Public Health, Boston, MA, USA
5. Richard A. and Susan F. Smith Institute, Beth Israel Deaconess Medical Center, Boston, MA, USA
6. Department of Biomedical Informatics, Harvard Medical School, Boston, MA, USA
7. Kempner Institute for the Study of Natural and Artificial Intelligence, Harvard University, Allston, MA, USA

*** Correspondence:**

Andrew R Weckstein
aweckstein@g.harvard.edu
677 Huntington Avenue, Boston, MA 02115
Kresge 911

**Work in progress:** This manuscript reports ongoing research and has not yet undergone peer review. Comments and feedback are welcome.

## ABSTRACT

Tabular foundation models based on prior-data fitted networks (PFNs) achieve strong predictive performance, but their value for causal inference is unclear. We evaluated whether PFNs, which predict through in-context learning, could match or outperform conventional nuisance learners in supporting accurate causal estimation and reliable uncertainty quantification (inference). Learners modeled propensity scores and outcome regressions within doubly robust estimators of average causal effects, with and without cross-fitting, across synthetic and semi-synthetic settings with varying confounding complexity and effect heterogeneity.

Under simple parametric data-generating processes, the PFNs were competitive with correctly specified parametric regressions and ensembles containing them. In more complex settings, they outperformed all conventional learners evaluated, including richly specified Super Learner ensembles, for both accuracy and uncertainty quantification. As sample size increased, only the two PFN models (TabPFN and Mitra) consistently approached nominal confidence interval coverage, whereas under-coverage persisted for conventional learners. In some settings, PFNs achieved near-nominal coverage and well-calibrated standard errors even without cross-fitting. At the smallest sample size ($n = 200$) and in the higher-dimensional semi-synthetic settings, PFNs and other flexible learners exhibited poor coverage without cross-fitting. For PFNs, this appeared to be driven by propensity score overfitting consistent with in-sample memorization of treatment labels. These findings highlight the promise of prior-data fitted networks as in-context nuisance learners for causal estimation, while underscoring the need for further evaluation in challenging settings.

## INTRODUCTION

Deep learning methods have reshaped prediction across scientific disciplines,[1–7] yet have historically struggled to outperform other established prediction methods on tabular data.[8–10] An emerging class of tabular foundation models based on prior-data fitted networks (PFNs) has challenged this pattern, demonstrating competitive predictive accuracy across a range of benchmarks.[11–14] Whether these predictive gains translate into improved causal effect estimation, however, remains unclear.

Estimation of causal effects often depends on accurate modeling of so-called “nuisance” functions. For example, estimation of the average treatment effect (ATE) for a point-treatment setting may require modeling the treatment (propensity score) and/or outcome functions. Flexible machine learning (ML) algorithms offer an appealing approach to nuisance estimation because they can adapt to nonlinearities and complex interactions that are difficult to pre-specify in parametric models, since the true data-generating process is unknown. However, greater flexibility introduces practical and statistical tradeoffs, including sensitivity to overfitting and challenges for valid uncertainty quantification.[15–17]

Modern statistics has therefore developed complementary strategies to enable flexible nuisance estimation while preserving robustness and inference.[16] First, ensemble algorithms, such as the SuperLearner,[18] hedge against uncertainty about the true data-generating process by combining predictions from a diverse library of base learners. Second, doubly robust estimators provide partial protection against misspecification and support valid inference under weaker convergence requirements.[19,20] Third, cross-fitting (sample splitting) relaxes restrictions on the complexity of nuisance learners, permitting a wider range of adaptive ML algorithms to be used within doubly robust frameworks.[15,21]

These practices have primarily been evaluated for conventional fit-then-predict algorithms trained directly on the analysis dataset, such as parametric regressions or tree-based methods. Modern PFN-based foundation models differ in two important ways. First, they are *pre-trained*: a transformer-based neural network[22] is trained once, in advance, on millions of synthetic datasets drawn from a broad prior over structural causal models, which encode diverse data-generating mechanisms.[13] Second, they predict through *in-context learning*: the model takes labeled observations as context and generates predictions for new inputs in a single forward pass, with no dataset-specific weight updates.[23,24] Figure 1 illustrates pre-training and in-context learning for TabPFN, one example from this model class.

Beyond their out-of-the-box convenience, the varied modeling strategies encoded during pre-training may improve PFNs' robustness across settings, serving a similar purpose to the diversity of learners in an ensemble. However, whether a PFN-based foundation model can match or outperform ensembles purpose-built for causal estimation is unknown. The role of cross-fitting for in-context learners with fixed weights at inference is also not well characterized. Motivated by its strong performance on tabular prediction tasks, we evaluate a pre-trained foundation model (TabPFN) as a nuisance learner for causal effect estimation.

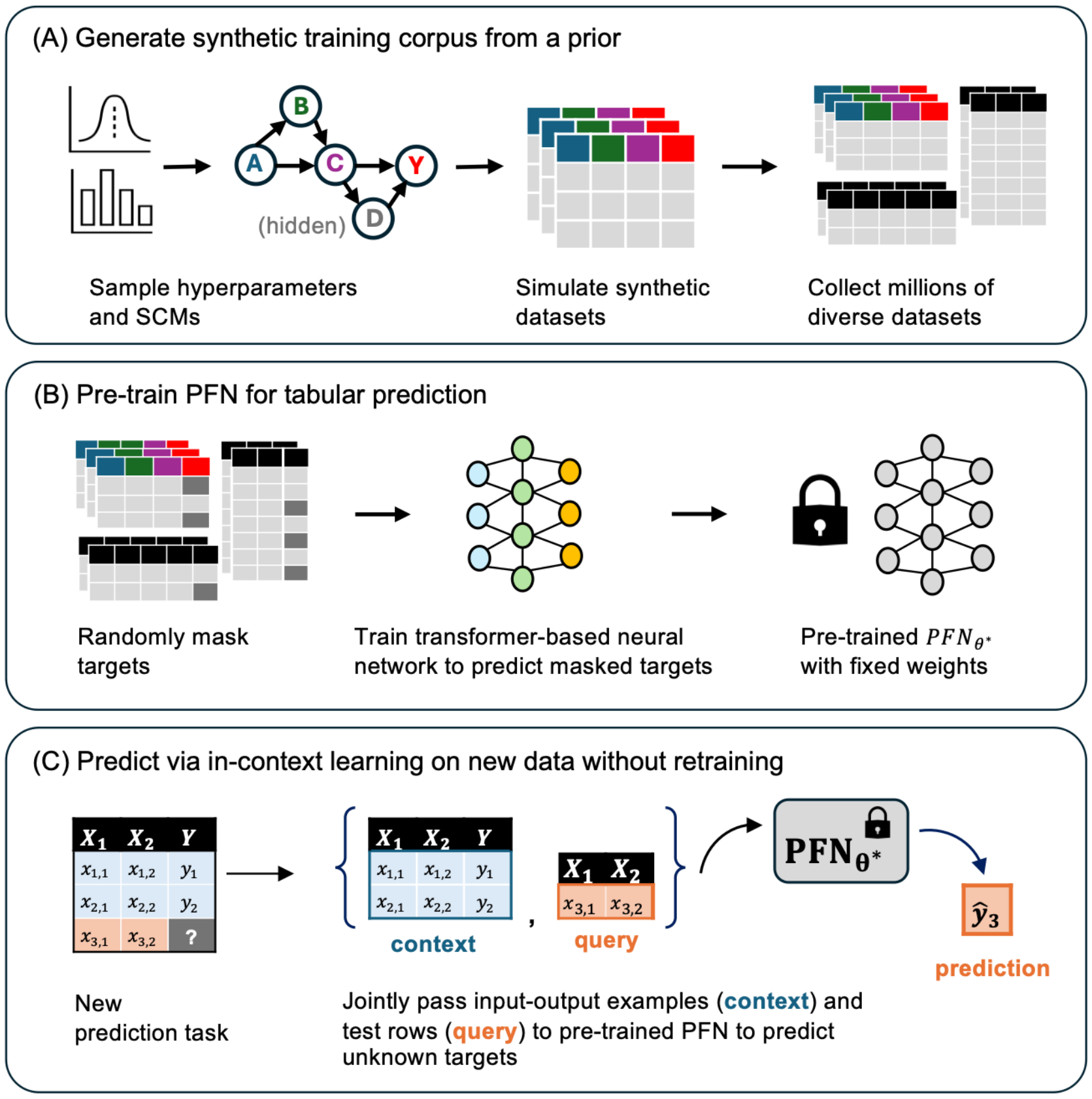

**Figure 1. Overview of pre-training and in-context learning for a tabular foundation model with prior-data fitted networks (TabPFN).**
(A) Sample hyperparameters and structural causal models (SCMs) from a prior over diverse data-generating processes, then simulate millions of tabular datasets.
(B) Train PFN to predict masked targets conditional on labeled context observations. Network weights are updated during pre-training and then fixed, yielding the pre-trained $\mathrm{PFN}_{\theta^*}$.
(C) Given supplied examples (blue) and query inputs (orange) from a new dataset, the pre-trained $\mathrm{PFN}_{\theta^*}$ predicts query targets in a single forward pass, without further training or weight updates.

# METHODS

We conducted two simulation studies. **Simulation 1** (synthetic) investigated tradeoffs with use of parametric versus flexible ML learners for estimation of average treatment effects under varying degrees of confounding complexity. **Simulation 2** (semi-synthetic/plasmode) assessed robustness across a broad set of realistic data-generating processes (DGPs). Across both studies, we compared TabPFN to nuisance learners spanning parametric regression, standalone ML methods, and Super Learner ensembles, each implemented with and without cross-fitting.

We first give a brief conceptual overview of identification, estimation, and why the choice of nuisance learner matters for both bias and valid inference. We then describe the design of Simulations 1 and 2.

## Causal estimand and identification

Let $(X_i, A_i, Y_i)$ be a sample of independent and identically distributed observations for individuals $i = \{1, \ldots, n\}$, where $X$ is a vector of baseline covariates, $A$ is a binary treatment taking values 0 (untreated) or 1 (treated), and $Y$ is a continuous outcome. We use $Y_i^a$ to represent the potential (i.e., counterfactual) outcome[25] that would be observed if treatment $A_i$ were set to value $a$.

Our primary causal estimand is the average treatment effect (ATE), defined as the expectation of the difference of potential outcomes had everyone received treatment versus no treatment:

$$ATE\colon \psi = E[Y^{a=1} - Y^{a=0}]$$

The ATE is identifiable from the observed data under three conditions:

I. Conditional exchangeability: $Y^a \perp A|X, \text{ for } a \in \{0,1\}$
II. Positivity: $0 < \Pr(A = 1|X = x) < 1$, for all covariate values in the support of X.
III. Consistency: If $A_i = a$, then $Y_i^a = Y_i$, for every $a \in \{0,1\}$.

Conditional exchangeability and positivity conditions hold by design across our simulations, and we assume consistency throughout. Under these conditions, the ATE is identified by the following equivalent functionals of the observed-data distribution:

$$\psi = E[\mu_1(X) - \mu_0(X)] \quad (1)$$

and

$$\psi = E\left[\frac{AY}{\pi(X)} - \frac{(1-A)Y}{1-\pi(X)}\right] \quad (2)$$

Where $\mu_a(x) = E[Y|A = a, X = x]$ denotes the true conditional outcome means, and $\pi(x) = \Pr(A = 1|X = x)$ the true propensity score.

## Estimation

All ATE estimators considered here rely on modeling the propensity score and/or conditional outcome nuisance models $\left(\pi(x),\ \mu_a(x)\right)$ with the observed data. For example, replacing the unknown nuisance functions with estimates and population expectations with sample averages in Equations 1 and 2 yields g-computation and inverse probability weighting (IPW) estimators, respectively. We implemented these estimators as sensitivity analyses (Suppl. S1). For our primary analyses, we use two doubly robust estimators of $\psi$: augmented inverse probability weighting (AIPW)[15,20,19] and targeted minimum loss-based estimation (TMLE).[26,27] Both AIPW and TMLE combine predictions from the propensity score and outcome models such that, under suitable regularity conditions, they remain consistent for the ATE if either model is consistently estimated.[19] Standard errors and 95% confidence intervals for AIPW and TMLE were obtained using the variance of the estimated influence curve (Suppl. S1).

## Nuisance estimation and statistical inference

The choice of nuisance learner affects both estimation and inference for $\psi$. We summarize the main theoretical considerations informally and refer to previous reviews[28,29] for precise conditions and alternative approaches. Obtaining valid Wald-type confidence intervals for $\psi$ with asymptotically correct coverage requires that the ATE estimation error, standardized by its estimated standard error, converges to a standard normal distribution. This in turn imposes conditions on the underlying nuisance models, so that the additional error introduced by nuisance estimation is negligible relative to the ATE estimator's sampling variability. Two important considerations are whether nuisance estimates become sufficiently accurate as sample size increases (sufficiently fast convergence rates), and whether the consequences of data-adaptive fitting are adequately controlled (empirical-process control[29]). This latter requirement can be achieved either by restricting learner complexity or via sample splitting.[28]

Under standard regularity conditions, correctly specified parametric regressions with a fixed number of parameters generally satisfy both requirements: nuisance estimation errors decrease at the $n^{-1/2}$ rate, satisfying the convergence-rate requirement, and their restricted model complexity provides the required empirical-process control. But parametric models impose strong assumptions on the structure of the true nuisance functions, leaving them susceptible to misspecification; when misspecified, they may converge to incorrect functions, potentially biasing ATE estimates and invalidating inference. In contrast, ML algorithms can reduce misspecification bias by flexibly modeling $\pi(x)$ and $\mu_a(x)$ with fewer functional form assumptions. However, this flexibility can complicate inference in two ways.

First, highly flexible learners may require more data to achieve a given level of accuracy, sometimes converging too slowly to support standard asymptotic approximations,[15,30] particularly when many covariates are involved (the "curse of dimensionality"). Our primary AIPW and TMLE estimators help address this challenge through Neyman orthogonality, which reduces first-order sensitivity of the ATE estimator to nuisance errors. This permits valid inference with a larger class of slower-converging ML learners, provided that the product of propensity score and outcome model estimation errors still decreases faster than $n^{-1/2}$ and other regularity conditions hold.[15,28] Second, flexible learners are prone to overfitting, which can distort inference when the same observations are used for nuisance fitting and ATE estimation. Complexity restrictions, such as Donsker conditions, can control these effects but may not hold for highly adaptive ML learners.[15,30] Alternatively, cross-fitting,[15,21,31] a form of sample splitting,[32] sidesteps the need for Donsker-type conditions by using out-of-fold nuisance predictions, so that each unit's predictions come from models fit without that unit's data. We therefore evaluated each nuisance learner with and without cross-fitting.

## Simulation 1: Fully synthetic data with simple and complex confounding

### Set-up for simulation 1

The first fully synthetic simulation used the simple (parametric) and complex confounding designs described by Naimi et al,[16] adapted from Kang & Schafer.[33] In both the simple and complex scenarios, the binary treatment $A$ and continuous outcome $Y$ were generated from parametric functions of four independent standard normal confounders $\{X_1, X_2, X_3, X_4\} \overset{iid}{\sim} \mathcal{N}(0,1)$. The true treatment effect was fixed across individuals ($\psi = 6,$ standardized effect size $d = 0.71$). We generated 200 Monte Carlo replicates for each of three sample sizes ($n = 200, 1200, 5000$).

In the simple scenario, models were fit using the observed $\{X, A, Y\}$ values from each simulated dataset. Therefore, the true nuisances were simple parametric functions of the observed data. In the complex scenario, only nonlinearly transformed versions of the covariates $Z_1, \ldots, Z_4$ were available for analysis (see Suppl. S1). This structure makes nuisance modeling more challenging because it requires learning complex non-linear relationships between $Z$ and $(A, Y)$, increasing the risk of misspecification.

**Learners for nuisance functions**

We explored a range of nuisance learning algorithms (see Table S1 for details). Parametric methods included generalized linear models (GLM), with logistic regression for $\pi(x)$ and linear regression for $\mu_a(x)$. These models will be correctly specified in simulation 1's simple scenario and misspecified in the complex scenario. The primary individual ML learners included: Random Forest as a standard adaptive learner benchmark,[34] the highly adaptive lasso (HAL) as a more recently proposed semi-parametric method with strong theoretical guarantees,[35,36] and TabPFN (v3) as the core tabular foundation model.[13,37] In contrast to the Random Forest and HAL learners, which received some hyperparameter tuning via cross-validation, the TabPFN learner was used out-of-the-box with default settings. Several other ML learners were implemented as secondary analyses for simulation 1, including a different tabular foundation model from the PFN class (Mitra).[38]

**Ensembles**

We used three Super Learner (SL) ensemble configurations,[18] implemented as convex stacking algorithms, with base learner weights determined by K-fold cross-validation. To provide a demanding comparison for TabPFN, two of these ensembles ("SL Flexible" and "SL Lightweight") were chosen because they were the top-performing algorithms in previous studies describing the same simulation 1 setting.[16,30] The "SL Flexible" ensemble included ten base learners from three flexible algorithm families: two Random Forest configurations, two XGBoost models, and six variants of generalized additive models (GAM), each considering main terms and all 2-way interactions.[16] The "SL Lightweight" ensemble library consisted of parametric GLMs, stepwise regression with pairwise interactions, multivariate adaptive regression splines, and the simple mean.[30] Finally, the third "SL Default" ensemble used the default base learners from the *tmle* and *SuperLearner* R software packages,[39,40] including GLM, BART, GAM, and LASSO. Because SL Lightweight and SL Default include a GLM base learner, they should asymptotically match the correctly specified parametric regression in simulation 1's simple scenario.[18]

**Evaluation metrics**

The primary evaluation metrics included bias (absolute relative bias), root mean squared error (RMSE), 95% confidence interval coverage, and run time. Relative bias was computed as the average per-replicate difference between the estimated and ground truth ATE, expressed as a percentage of the true causal effect. We report RMSE in absolute terms and as a percentage of the oracle RMSE. Coverage was defined as the proportion of Wald-type confidence intervals that included the true causal effect. We also calculated SE calibration ratios as the ratio of the average model-reported SE to the empirical SD of the point estimate across simulations.

## Simulation 2: Plasmode simulations across diverse data generating processes (DGPs)

Simulation 2 evaluated robustness across many different DGP structures, grounded in real covariate data from the Collaborative Perinatal Project, a longitudinal study of pregnancy outcomes.[41] We used the plasmode-type simulation design from the 2016 Atlantic Causal Inference Conference (ACIC) data analysis competition to generate treatments and outcomes from the fixed covariate matrix.[42] We simulated 2,200 different datasets across 44 distinct DGPs by varying different “knob” settings, including the degree of nonlinearity for treatment and outcome models, treatment prevalence, magnitude of confounding, degree of treatment effect heterogeneity, and full or partial propensity score overlap (Table S2). For each DGP, we simulated 50 Monte Carlo replicates, each restricted to a sub-sample of 1,000 observations drawn without replacement from the original 4,802 mother-child pairs. Bias and RMSE were evaluated against replicate-specific subsample effects, whereas coverage and SE calibration were evaluated against corresponding full-cohort effects (See Suppl. A6). Simulation 2 targeted the sample ATE and the sample average treatment effect on the treated (ATT), $\delta = E[Y^{a=1} - Y^{a=0}|A = 1]$. The conditions for point-identification of the ATT are slightly weaker than those for the ATE, since conditional exchangeability requires only $Y^0 \perp A|X$ and positivity requires $Pr(A = 1|X = x) < 1$. While the ATE used only settings with full propensity score overlap to ensure identifiability (35 DGPs), the ATT was evaluated over all 44 DGPs (see Suppl. S1).

Simulation 2 used the same primary nuisance learners as simulation 1. To accommodate the higher-dimensional setting ($p = 80$ covariate levels), some ensemble base learners used screening algorithms to pre-select a subset of predictors within each run (Table S1).

Analyses were executed in R (version 4.5.2) and Python (version 3.11), and our study code is available on GitHub: Tabular-FM-causal-simulations, ACIC-plasmode-pipeline. The supplementary materials contain further methods detail (Suppl. S1) and extended results (Suppl. S2).

# RESULTS

## Simulation 1 Results

Figure 2 summarizes performance in simulation 1 with the TMLE estimator and cross-fitting, averaged across the three sample sizes. ATE bias (2A), RMSE (2B), and confidence interval coverage (2C) are plotted for the simple versus complex scenario.

Most learners faced a tradeoff in relative performance across the two scenarios. As expected, the parametric learner performed best in the simple scenario, where treatment and outcomes were simulated with basic parametric functions of the observed covariates, since models were correctly specified by construction (average bias 0.3%, coverage 96%). Ensembles with parametric base learners (SL Lightweight and SL Default) and GAM showed similarly low bias/RMSE and at least nominal (≥95%) coverage on average for the simple scenario. However, these same learners performed poorly in the complex scenario (bias 18-30%, coverage 25-50%). The SL Flexible ensemble with more adaptive base learners showed a reverse pattern: it was relatively weaker than other learners in the simple scenario (average bias 6%, coverage 92%), but was the strongest of the non-PFN learners in the complex scenario (average bias 15%, coverage 66%).

TabPFN and the other PFN-based foundation model variants were the only nuisance learners to perform near the top in both settings. In the simple scenario, the PFNs came close to correctly specified parametric benchmarks (bias 1-2%, coverage 95-96%). In the complex scenario, they had the lowest average bias and highest average coverage by a wide margin (bias <10%, coverage 90-91%).

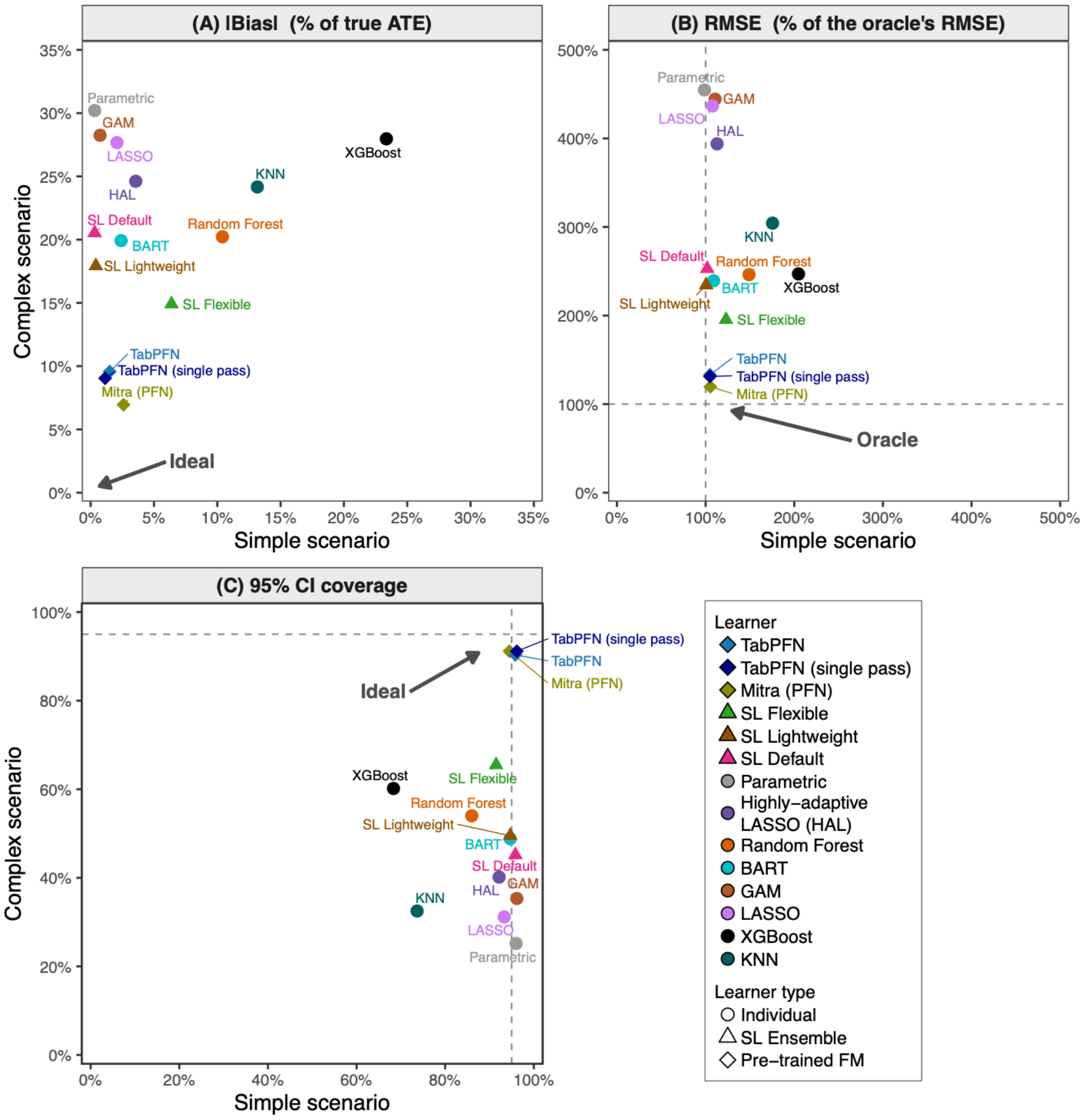

**Figure 2. TMLE estimation of the average treatment effect with cross-fitted nuisances in synthetic simulation 1, averaged across sample sizes.** Points summarize each nuisance learner's performance in the simple versus complex scenarios, averaged across sample size, for (A) absolute bias as percentage of true ATE, (B) RMSE as percentage of corresponding oracle RMSE, and (C) coverage of 95% confidence intervals. Metrics were calculated separately from 200 replicates at each sample size before averaging. The oracle model runs TMLE with the true nuisance functions in place of fitted ones.

*Abbreviations: ATE = average treatment effect, CI = confidence interval, RMSE = root mean squared error, SL = super learner ensemble, TMLE = targeted minimum loss-based estimation.*

The cross-fit averages in Figure 2 mask important variations by sample size, as illustrated in Figure 3, where each cell displays the average from 200 simulation replicates within a specific configuration (Table 1 shows detailed results for the $n = 1200$ sample size). For instance, in the simple scenario at $n = 200$ without cross-fitting, bias/RMSE increased across all non-parametric learners, and coverage dropped to 87–88% for the parametric-based ensembles and to 66% for TabPFN. In the complex scenario, shown in the rightside of each Figure 3 panel, TabPFN's advantage grew with increasing $n$. At the largest $n = 5000$ sample size, TabPFN bias was <2% both with and without cross-fitting, compared to 8-10% for the three SL ensembles and ~28-30% for HAL and parametric (Figure 3A). While the PFNs' bias in the complex scenario increased at the smallest $n = 200$ sample size, this was a consistent pattern across all flexible non-parametric learners, with an inverse relationship between bias/RMSE and $n$.

The PFN models were also the only learners - among all ensembles and standalone algorithms - to ever achieve near-nominal coverage in the complex scenario (Figures 2-3, Table 1). For other learners, neither larger sample size nor cross-fitting could rescue coverage. Indeed, for non-PFN learners, coverage tended to *worsen* with increasing sample size, indicating that bias declined more slowly than the estimator's sampling error.

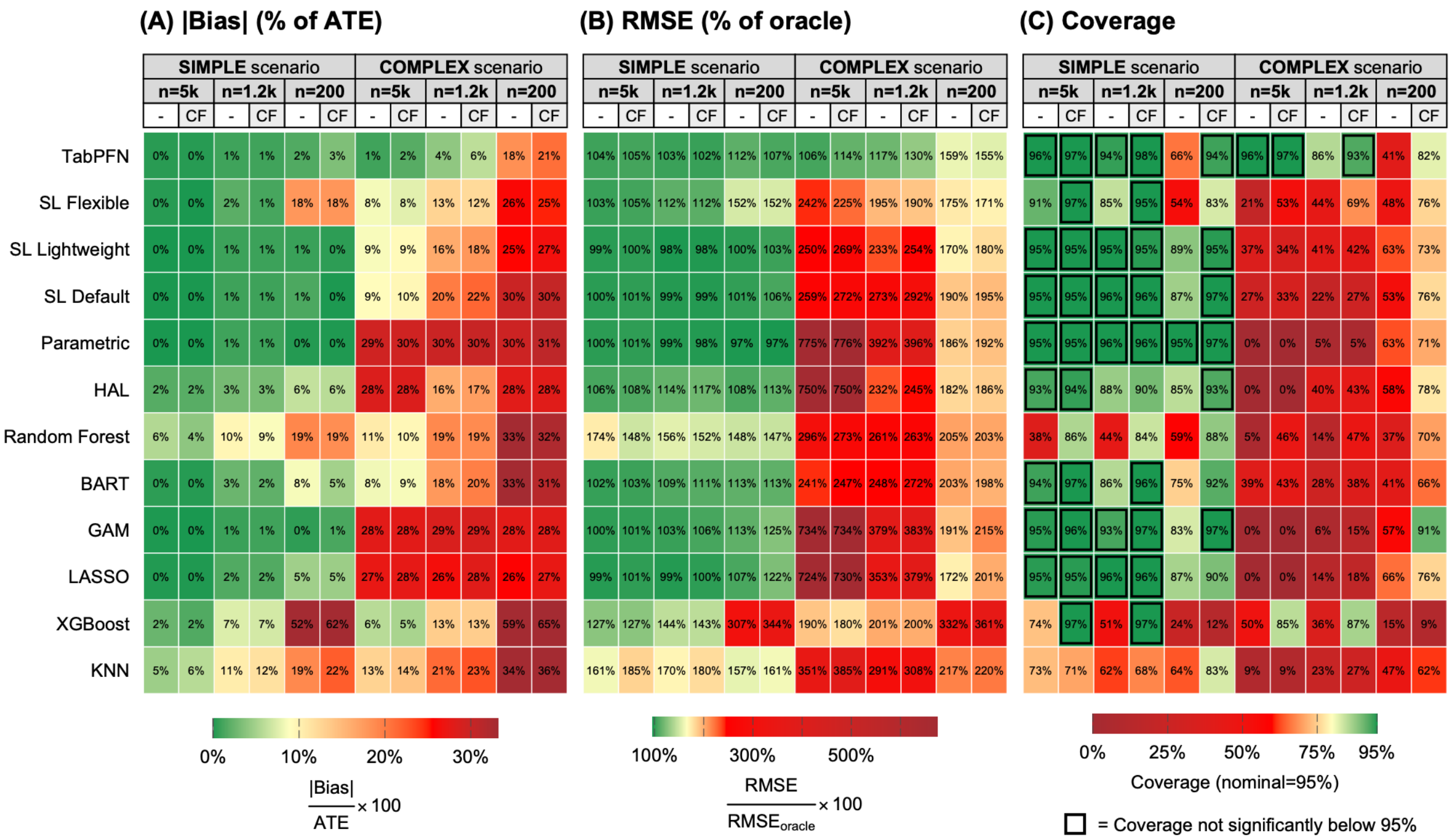

**Figure 3. TMLE estimation of the average treatment effect in synthetic simulation 1 by scenario, sample size, and cross-fitting.** Each cell summarizes nuisance learner performance across 200 simulation replicates within a given configuration (scenario × sample size × cross-fitting), including: (A) absolute bias as percentage of the true ATE, (B) RMSE as percentage of oracle RMSE, and (C) coverage of 95% confidence intervals. Rows represent nuisance learners and columns represent simulation configurations. Color scales are capped at each metric's 95th-percentile value. The oracle model runs TMLE with the true nuisance functions in place of fitted ones; fitted nuisances may occasionally attain lower RMSE than the oracle in finite samples. Bordered cells in panel (C) indicate coverage not significantly below nominal (95%), via a one-sided z-test using the Monte Carlo standard error under the null. *Abbreviations:* ATE = *average treatment effect*, *CF = cross-fit, HAL = highly-adaptive LASSO, RMSE = root mean squared error, SL = super learner ensemble, TMLE = targeted minimum loss-based estimation.*

**Table 1.** Nuisance learner performance for TMLE estimation of the average treatment effect in synthetic simulation 1 at the moderate sample size (n=1200)

| | | **SIMPLE SCENARIO** *(not cross-fit / cross-fit)* | | | | | **COMPLEX SCENARIO** *(not cross-fit / cross-fit)* | | | | | |
|---|---|---|---|---|---|---|---|---|---|---|---|---|
| **Learner type** | **Learner** | **\|Bias\|** (% of ATE) | **RMSE** | **SE ratio**[a] | **CI width** | **Coverage**[b] | **\|Bias\|** (% of ATE) | **RMSE** | **SE ratio**[a] | **CI width** | **Coverage**[b] | **Run time**[c] (sec) |
| **PFNs** | TabPFN | 0.9% / 0.9% | 0.49 / 0.49 | 0.95 / 1.14 | 1.80 / 2.15 | **94%** / **98%** | 4.0% / 6.0% | 0.56 / 0.61 | 0.82 / 1.12 | 1.62 / 2.20 | 86% / **93%** | 2.3s / 7.8s (CPU 11s / 30s) |
| | TabPFN (single pass)[d] | 0.8% / 1.0% | 0.48 / 0.48 | 1.00 / 1.16 | 1.87 / 2.17 | **96% /** **98%** | 3.7% / 5.7% | 0.55 / 0.61 | 0.83 / 1.10 | 1.63 / 2.17 | 85% / **93%** | 0.4s / 1.2s (CPU 1.5s / 3.7s) |
| | Mitra | 1.1% / 1.1% | 0.47 / 0.47 | 0.94 / 1.02 | 1.72 / 1.89 | **95% /** **96%** | 3.0% / 3.2% | 0.51 / 0.51 | 0.86 / 1.04 | 1.63 / 1.95 | 91% / **94%** | 1.3s / 6.1s (CPU 28s / 71s) |
| **Ensembles** | SL Flexible | 1.9% / 1.2% | 0.53 / 0.53 | 0.72 / 0.98 | 1.47 / 2.01 | 85% / **95%** | 12.7% / 12.2% | 0.92 / 0.90 | 0.63 / 0.97 | 1.29 / 2.02 | 44% / 69% | 44s / 210s |
| | SL Lightweight | 0.8% / 0.8% | 0.46 / 0.46 | 0.96 / 1.01 | 1.74 / 1.84 | **95%** / **95%** | 16.3% / 17.9% | 1.10 / 1.20 | 0.83 / 0.92 | 1.69 / 1.95 | 41% / 42% | 0.5s / 2.5s |
| | SL Default | 0.9% / 0.9% | 0.47 / 0.47 | 0.95 / 1.04 | 1.73 / 1.90 | **96% /** **96%** | 20.0% / 21.6% | 1.29 / 1.38 | 0.85 / 1.03 | 1.57 / 1.96 | 22% / 27% | 14s / 57s |
| **Conventional standalone** | Parametric | 0.9% / 0.9% | 0.47 / 0.47 | 1.00 / 1.03 | 1.82 / 1.88 | **96%** / **96%** | 29.9% / 30.2% | 1.86 / 1.87 | 1.01 / 1.03 | 1.88 / 1.95 | 5% / 5% | 0.01s / 0.1s |
| | HAL | 2.9% / 3.0% | 0.54 / 0.56 | 0.83 / 0.86 | 1.67 / 1.78 | 88% / 90% | 16.1% / 17.3% | 1.10 / 1.16 | 0.75 / 0.93 | 1.55 / 1.90 | 40% / 43% | 77s / 99s |
| | Random Forest | 9.7% / 8.6% | 0.74 / 0.72 | 0.62 / 1.05 | 1.10 / 2.04 | 44% / 84% | 19.0% / 18.9% | 1.24 / 1.24 | 0.62 / 1.08 | 1.15 / 2.17 | 14% / 47% | 0.7s / 3.4s |
| | Oracle[e] | 0.9% | 0.47 | 0.98 | 1.82 | **94%** | 0.9% | 0.47 | 0.98 | 1.82 | **94%** | - |

Abbreviations: *HAL = highly-adaptive LASSO, PFN = prior-data fitted network, TMLE = targeted minimum loss-based estimation, RMSE = root mean squared error, SL = super learner ensemble.*

**Note.** Evaluation metrics for TMLE estimation of the average treatment effect (ATE) across 200 simulations of the n=1200 sample size for simulation 1. Each cell shows values for *without cross-fitting / with cross-fitting*. We used the v3 weights for TabPFN and the v2 weights for Mitra.

[a] SE ratio (calibration) is the average model-reported SE from the efficient influence curve divided by the empirical standard deviation of point estimates across 200 simulations: 1 indicates perfect calibration, <1 overconfident, >1 conservative.

[b] Coverage values in bold are statistically indistinguishable from the Monte Carlo error of a correctly calibrated interval with nominal coverage (95%), via a one-sided z-test.

[c] Run times were measured as the average wall time to fit both propensity score and outcome nuisance fits across 35 simulation replicates of the complex scenario. CPU times used a local laptop with 12 CPU cores and 24 GB memory; GPU times used an A100 GPU.

[d] TabPFN (single pass): one forward pass per prediction, instead of TabPFN's default average over 8 passes.

[e] Oracle model runs TMLE with true nuisance functions in place of fitted ones. Fitted nuisances can still beat the oracle in finite samples. There is no cross-fit equivalent for oracle results.

**Cross-fitting and calibration**
Among the non-PFN ML learners, model-based SEs were systematically overconfident without cross-fitting (SE calibration ratio<1), yielding confidence intervals that were too narrow (Table 1). Cross-fitting markedly improved SE calibration and coverage. The extent of under-coverage, and the corresponding benefit of cross-fitting, increased with learner adaptivity: highly adaptive tree-based learners and SL ensembles that included them benefitted most, whereas more constrained learners (e.g., HAL, SL Lightweight) showed smaller gains.

Because the PFNs tended to be better calibrated than other non-parametric learners at n=1200/5000, cross-fitting sometimes worsened calibration by making SEs conservative and intervals wider than necessary (Tables 1, S3-S5). For example, in the n=1200 simple scenario, cross-fitting degraded TMLE-TabPFN's SE calibration (SE ratio 0.95 → 1.14). In contrast, at n = 200, TabPFN was the worst-calibrated learner without cross-fitting (SE ratio <0.5) and gained the most from cross-fitting (coverage +27.5 and +41.5 percentage points in the simple and complex settings). This appeared to be driven by overfitting of the propensity score model: $\hat{\pi}(x)$ AUCs were 0.95-0.97 in-sample (no cross-fitting) versus 0.70-0.72 out-of-fold (with cross-fitting) (Table S6). Other flexible learners showed similar behavior, but for the PFNs the issue diminished rapidly with increasing sample size.

**Investigating other model variants**
Results were similar for a streamlined TabPFN variant without internal test-time averaging (TabPFN single pass) and for Mitra[38], a different PFN-based foundation model (Table 1, Suppl. S2). Adding TabPFN to a lightweight SL ensemble improved coverage without cross-fitting at $n = 200$, with no meaningful difference after cross-fitting (Table S4).

**Runtime**
TabPFN's cross-fit run time (30s CPU, 7.8s GPU) fell between SL Lightweight (2.5s) and the other SL ensembles (57-210s). Run time dropped to 1.2s (GPU) for the streamlined TabPFN (single pass) variant (Tables 1, S3).

**Comparing estimators**
Findings were generally concordant between the two doubly robust estimators, except AIPW-TabPFN had elevated bias in the non-cross-fit simple-scenario at n=200 (13.1% AIPW vs 1.5% TMLE), though this discrepancy was eliminated with cross-fitting (Figure S1). Relative to the doubly robust estimators, IPW and G-computation had higher ATE bias/RMSE across most cells in both scenarios.

**Covariate balance assessment**
Because the true DGP is usually unknown in practice, analysts often use covariate balance diagnostics to assess confounding control and potential model misspecification. We computed absolute standardized mean differences (ASMDs) after inverse probability weighting with each learner's estimated propensity scores, for the complex scenario at $n = 1200$. We assessed balance separately for the latent "true" generating confounders $\{X_1, X_2, X_3, X_4\}$ (Figure 4A) and the observed, model-provided covariates $\{Z_1, Z_2, Z_3, Z_4\}$, which were nonlinear transformations of the generating confounders (Figure 4B). In Figure 4A, TabPFN achieved substantially lower ASMDs on the generating confounders and lower IPW ATE bias than the other learners, with and without cross-fitting. Balance on the latent generating confounders was strongly aligned with ATE bias across learners. In contrast, balance on the observed transformed covariates $Z$ - representing the naïve diagnostic an analyst would see without considering the true functional forms - was a weak proxy for bias (Figure 4B). For example, parametric appeared well balanced on $Z$, despite poor balance on the true confounders (Figure 4A) and high ATE bias, consistent with misspecification of the underlying structure.

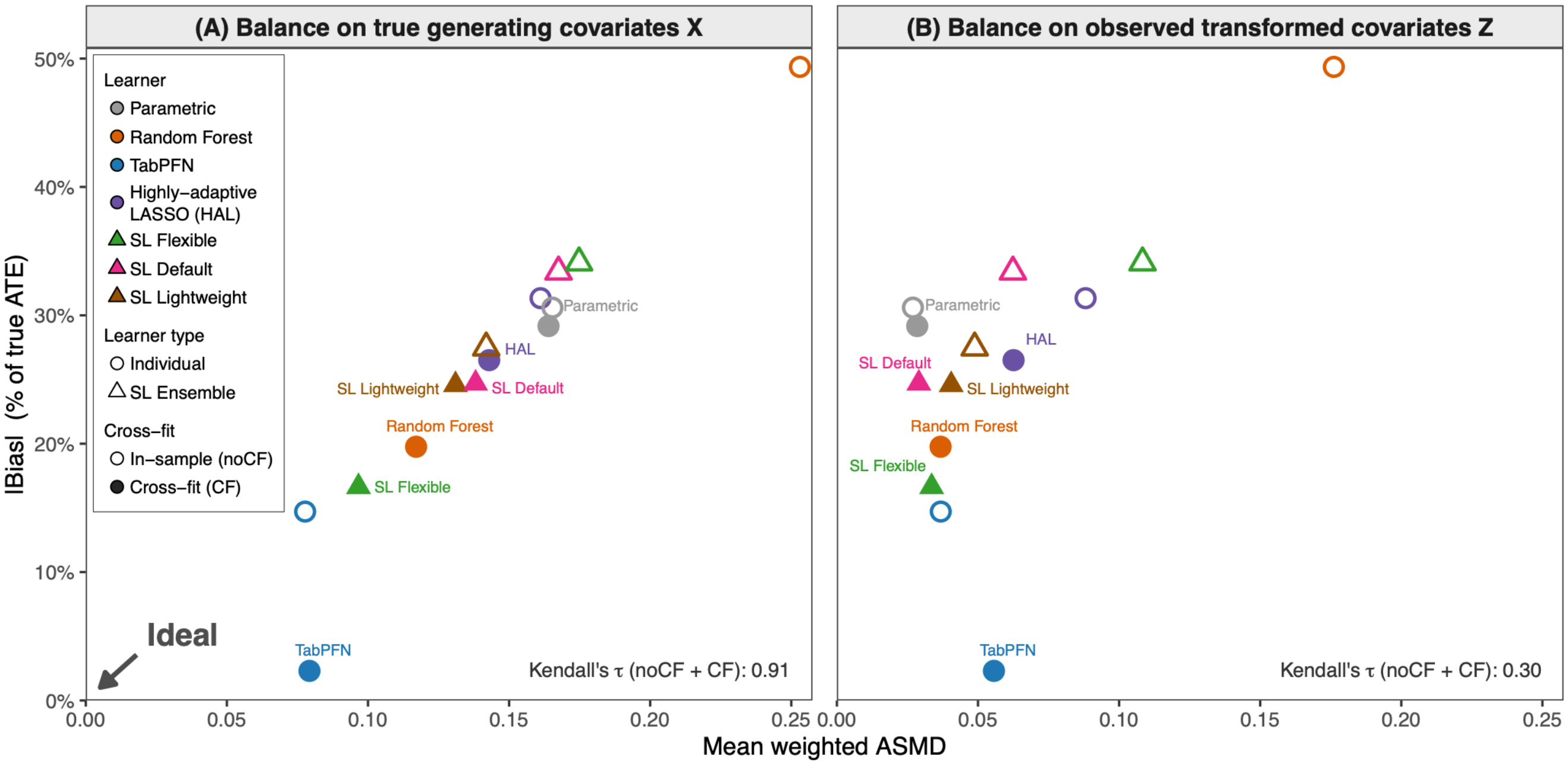

**Figure 4. Covariate balance and bias for IPW estimation of the average treatment effect in the complex scenario for synthetic simulation 1 (n=1200).** Plots the mean ASMD for the (A) true generating covariates $X$ and the (B) observed transformations $Z$, against relative bias of the IPW ATE estimator. ASMDs were averaged over the four covariates across 50 replicates of the $n = 1200$ complex scenario, where the data-generating process is governed by the latent confounders $X$. Lower values indicate better balance and lower absolute bias. Kendall's $\tau_b$ measures how well the ordering of learners by balance agrees with their ordering by IPW bias. *Abbreviations: ASMD = absolute standardized mean differences, IPW = inverse probability weighting.*

## Simulation 2 Results

### ATE and ATT Estimation with cross-fitting

Across all evaluated DGP settings, TabPFN was the best learner on average for cross-fit ATE and ATT estimation in simulation 2. For TMLE with cross-fitting, it obtained the lowest macro-average RMSE and nearly half the macro-average bias relative to the next best learner (relative bias 0.8% vs 1.6% for ATE, 0.9% vs 1.7% for ATT) (Figure 5). TabPFN's advantage was consistent across the range of explored DGP settings: For TMLE with cross-fitting, TabPFN had the lowest setting-specific mean RMSE for all ATE-identified DGPs (35/35, 100%) and for most ATT-identified DGPs (38/44, 86%). SL Lightweight and SL Default were the next best performers for ATE/ATT estimation, followed by SL Flexible, HAL, Random Forest, and Parametric. With TMLE and cross-fitting, coverage was not significantly below the nominal level in at least 94% of DGPs for TabPFN, at least 91% for SL Default, and at least 89% for parametric regression, compared with 3-70% of DGPs for the other learners, although parametric had confidence intervals more than four times wider than TabPFN's (Suppl. S2). Because of the sub-sampling procedure and use of only 50 replicates per DGP, simulation 2 coverage results should be interpreted with caution (see Suppl. S1).

**Comparing estimators and cross-fitting configurations for simulation 2**
AIPW and TMLE performed similarly for the ATE and ATT whenever nuisances were cross-fit and, without cross-fitting, for all non-TabPFN learners. The clear exception was non-cross-fit TabPFN: TMLE-TabPFN substantially underperformed AIPW-TabPFN for both estimands (ATE: bias 2.68% vs 0.84%, coverage 58% vs 85%; ATT: bias 6.65% vs 0.89%; coverage 59% vs 91%; Figures S3-S6). This discrepancy appeared to be driven by TabPFN's in-sample memorization of treatment values, the same overfitting observed in simulation 1. Without cross-fitting, TabPFN separated treated and untreated units almost perfectly (mean AUC≈1) despite weak treatment assignment in the true DGP, while recovering the true $\pi(x)$ worst of all learners (RMSE 0.377 vs 0.076-0.141) and pushing predicted propensities toward the extremes (Table S7). While the strong outcome model largely protected AIPW from TabPFN's in-sample $\hat{\pi}(x)$, high-leverage inverse-probability terms appeared to destabilize the TMLE targeting step, leading to degraded TMLE performance.[43] Cross-fitting brought TabPFN's $\pi(x)$ discrimination in line with the true DGP and restored it as the best-performing nuisance learner for both doubly robust estimators.

Across the four cross-fit ATE/ATT estimators, IPW generally performed worst, whereas G-computation matched the doubly robust estimators. Comparing the TMLE/AIPW/G-computation vs IPW estimators, the cross-fit TabPFN had mean ATE bias of 1%/1%/1% vs 7%, and the cross-fit SL Lightweight achieved mean bias of 2%/2%/2% vs 12%, with similar results for the ATT (Figures S3-S4).

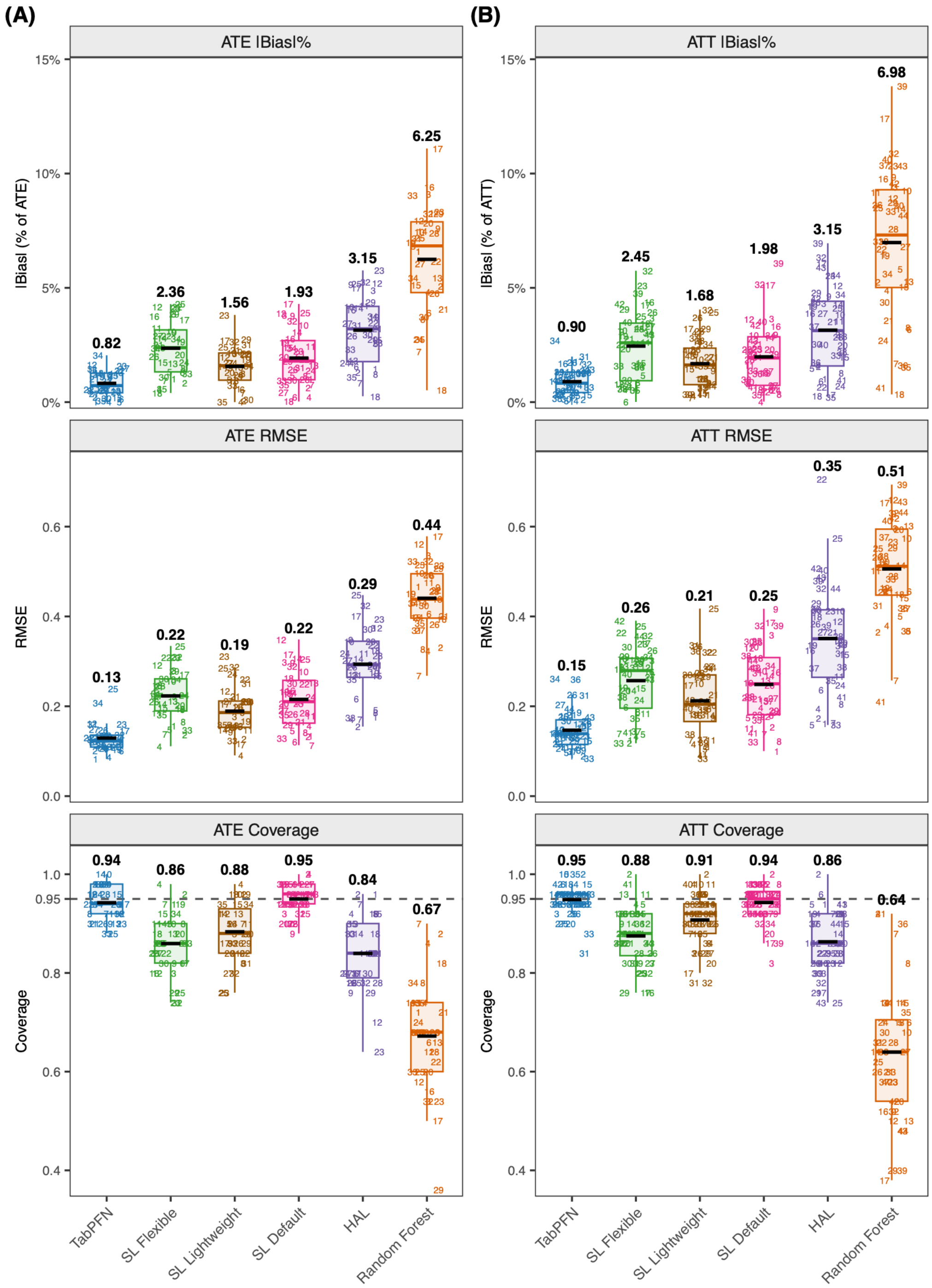

**Figure 5. TMLE estimation of the ATE and ATT with cross-fitted nuisances across diverse settings from plasmode simulation 2.**

Columns show (A) ATE estimation in 35 full-overlap settings and (B) ATT estimation in all 44 settings. Each point summarizes one nuisance learner's performance across 50 replicates within one data-generating process (DGP) setting; numbers identify DGPs listed in Table S2. Rows show absolute relative bias, RMSE, and coverage of 95% confidence intervals. Bias and RMSE use replicate-specific subsample targets whereas coverage uses full-cohort targets. Black bars and bold labels indicate unweighted macro-averages across all contributing DGP settings. Parametric regression is omitted for readability.

*Abbreviations: ATE = average treatment effect, ATT = average treatment effect on the treated, RMSE = root mean squared error, TMLE = targeted minimum loss-based estimation.*

## DISCUSSION

Because the true DGP, and therefore the optimal nuisance learner, is rarely known *a priori*, analysts using flexible ML for nuisance estimation must balance two practical considerations: whether flexibility actually improves over simpler parametric models when the underlying structure is complex, and how costly that flexibility is when the structure is simple. In our simulations, a pre-trained tabular foundation model (TabPFN) performed well across both regimes. In complex nonlinear confounding scenarios, TabPFN consistently outperformed richly specified ensembles, achieving low bias and approximately nominal coverage at all but the smallest sample size, with performance that was robust across DGP characteristics and causal estimands. In the simple parametric setting, TabPFN generally matched correctly specified parametric models and ensembles containing them. A supporting analysis with a second tabular foundation model pretrained on synthetic data (Mitra) showed a similar pattern for simulation 1, suggesting this robustness may extend to PFN-based models more broadly. No other (non-PFN) learner performed near the top in both settings. These findings extend prior evidence of state-of-the-art predictive performance to causal estimation and add to an emerging literature exploring PFNs for causal tasks.[13,14,44–47]

The recommended diverse ensembling approach[16,48] requires manual selection of base learners and screening algorithms, along with potentially expensive cross-validation and hyperparameter tuning. That untuned, pre-trained PFNs matched or outperformed purpose-built ensembles suggests that pre-training across a broad synthetic prior may offer an alternative way to hedge against DGP uncertainty. One possible explanation is that pre-training across diverse synthetic tasks provides flexibility analogous to combining multiple conventional learners, with model selection implicitly learned during pre-training and amortized at inference. This has obvious efficiency appeals, particularly given the competitive run times. However, PFN performance may depend on how well the pre-training prior represents the relevant data-generating structures. This implies both potential blind spots and avenues for improvement via modification of the training distribution.[38] For instance, PFN priors could be used to embed domain knowledge for specialized tasks, or to develop models for causal estimation in challenging regimes (e.g., unstable positivity, rare treatments, or small samples).

We examined TabPFN's behavior under the doubly robust and cross-fitting strategies recommended for flexible ML nuisance learners,[16] where its in-context-learning architecture might plausibly introduce special considerations. For all flexible learners, including TabPFN, doubly robust estimation with AIPW or TMLE remained superior to singly robust estimators across most scenarios. However TabPFN's accuracy advantage was most pronounced when modeling the continuous outcome, and in simulation 2 this relative strength made G-computation with TabPFN competitive with its doubly robust counterparts.

Cross-fitting generally improved SE calibration and coverage for flexible learners, consistent with theory and prior empirical findings.[15,21] For non-PFN learners, however, coverage in the complex simulation 1 scenario remained far below nominal even after cross-fitting, suggesting residual bias from misspecification that did not appreciably diminish with increasing $n$ relative to sampling variability. In contrast, the PFNs showed a distinctive pattern. At moderate-to-large sample sizes ($n = 1200/5000$) in simulation 1, TabPFN and Mitra had good SE calibration and low ATE bias in-sample, so cross-fitting yielded only moderate improvement and was sometimes counterproductive. At $n = 200$ in simulation 1, and for TMLE-TabPFN in simulation 2, the PFNs benefited substantially from cross-fitting. For both simulation 1 and 2, the PFNs' occasional failures were linked primarily to in-sample memorization of treatment labels, which cross-fitting appeared to mitigate. Whether this finite-sample pattern is specific to the settings we studied or reflects a broader limitation of these in-context learners for propensity score estimation in low-n/p regimes warrants further study. Other promising directions include uncertainty quantification derived directly from the Bayesian-inspired PFN predictive distribution,[13] and post-hoc calibration methods.[49]

Our study has several limitations. To ensure a ground-truth evaluation, we used simulations, which cannot capture the full diversity of real-world DGPs. However two design choices strengthen generalizability. First, we borrowed established simulation designs and their best-performing ensembles, creating a more demanding and objective test. Second, we evaluated a deliberately diverse range of settings, including synthetic and semi-synthetic designs, multiple sample sizes, simple and complex confounding structures, and two causal estimands. Further, the simulation 2 knobs allowed us to explore DGPs reflecting challenges commonly encountered in practice, including nonlinearity, strong confounding, heterogeneity, and limited overlap.

Even so, important boundaries remain. We evaluated only average causal effects, with continuous outcomes and binary treatments. Further, all scenarios guaranteed identifiability, with a sufficient adjustment set among the observed covariates.  Additional research is needed to understand empirical performance in settings where identifiability is not assured, and where potential covariates include mediators, colliders, or instruments.[50,51] Large foundation models may also be impractical with larger datasets and high-dimensional covariate settings, though model architecture is scaling rapidly.[37,52–54]

We also could not evaluate every possible ML algorithm or hyperparameter configuration. Notable methods warranting further evaluation include undersmoothing techniques[55], collaborative TMLE[56], pre-trained+fine-tuned models, other PFN variants[57–59], and even foundation models trained directly for in-context causal estimation.[47,60] We share the longstanding expectation that no single model class will dominate in every setting, which makes the robustness of TabPFN's performance across our simulations all the more notable.

In conclusion, we find that a pre-trained tabular foundation model offers practical advantages for causal nuisance estimation beyond conventional ML learners and hand-curated ensembles.

## CONFLICTS OF INTEREST

A.R.W., I.J.D., S.H.D., and M.Z. have no conflicts of interest to disclose. S.S. is co-principal investigator of an investigator-initiated grant to Brigham and Women's Hospital from Boehringer Ingelheim unrelated to the topic of this study. He is a consultant to Datavant unrelated to the topic of this study. His interests were declared, reviewed, and approved by Brigham and Women's Hospital and MGB HealthCare System in accordance with their institutional compliance policies.

## FUNDING

A.R.W. received training support from the National Institutes of Health (T32-HL098048).

## ACKNOWLEDGEMENTS

The authors wish to thank Professor Nima Hejazi (Department of Biostatistics, Harvard T.H. Chan School of Public Health) for his helpful guidance around implementation of the highly adaptive lasso, and Professor Jie Yang (Department of Medicine, Brigham and Women's Hospital and Harvard Medical School) for contributing feedback during the early stages of this project. Anthropic's Claude Opus (v5, v5.5) and Fable (v5) models were used to double-program and organize statistical code and for proofreading the authors' manuscript text. The authors take full responsibility for the submitted content.

## DATA AND CODE AVAILABILITY

Simulated datasets from the synthetic and plasmode settings can be reproduced using the provided code: https://github.com/a-weckstein/Tabular-FM-causal-simulations, https://github.com/a-weckstein/ACIC-plasmode-pipeline.

# APPENDIX

## TABLE OF CONTENTS

**Content**

*See accompanying excel file.

## A1. Estimands & Identifiability

We have i.i.d realizations $\{A_i, Y_i, X_i\}$ for individuals $i = 1, 2, \ldots, n$, where n is the number of individuals in the sample, $A$ is a binary treatment variable taking values 0 (untreated) and 1 (treated), $Y$ is a continuous outcome variable, and $X$ is a vector of baseline covariates. We will assume throughout that $X$ includes all confounders necessary for identification. We use notation $Y_i^a$ for counterfactual (potential) outcomes, representing the outcome that would be observed if treatment $A_i$ were set to value $a$. The following sections may omit individual subscripts for convenience.

### Average treatment effect (ATE)

$$ATE: \psi = E[Y^{a=1} - Y^{a=0}]$$

The average treatment effect (ATE) is the expectation of the difference in counterfactual outcomes had everyone received treatment versus no treatment. The ATE is point-identified from the observed data under three assumptions:

I. Exchangeability: $Y^{a=1} \perp A|X \;\; and \;\; Y^{a=0} \perp A|X$
II. Positivity across the entire covariate distribution: $0 < \Pr(A = 1|X = x) < 1, \text{for all x with positive density f(x)} > 0$
III. Consistency: For all individuals, if $A_i = a$, then $Y_i^a = Y_i$, for every $a \in \{0,1\}$, which implies non-interference between individuals' counterfactual outcomes and no hidden versions of treatment.

Under these conditions, the ATE is identified by $\psi = E[\mu_1(X) - \mu_0(X)]$, where $\mu_a(x) = E[Y|A = a, X = x]$ denotes the true conditional outcome means, and $\pi(x) = \Pr(A = 1|X = x)$ the true propensity score.

### Average treatment effect on the treated (ATT)

$$ATT: \delta = E[Y^{a=1} - Y^{a=0}|A = 1]$$

The average treatment effect on the treated (ATT) is the expectation of the difference in counterfactual outcomes for treatment versus no treatment, among the treated (A=1). Point-identifiability of the ATT requires:

I. Exchangeability: $Y^{a=0} \perp A|X$
II. Positivity across the *treated* covariate distribution, $\Pr(A = 1|X = x) < 1 \text{ for all x with f(x|A = 1)} > 0$
III. (Causal) Consistency: If $A_i = a$, then $Y_i^a = Y_i$, for every $a \in \{0,1\}$.

The positivity and exchangeability conditions here are weaker than those for the ATE. In particular, the weaker positivity condition is why the ATT is identified in the simulation 2 settings of partial ("one-term") overlap, whereas the ATE is not.

Under these conditions, the ATT is point-identified by $\delta = E[Y|A = 1] - E[\mu_0(X)|A = 1]$.

## A2. Estimators

We used four different estimators for each estimand. We focus here on each estimator's form for the ATE. Standard errors and 95% Wald-type confidence intervals (CIs) were obtained using the robust variance estimator for inverse probability weighting (IPW) and the variance of the efficient influence function for Augmented IPW (AIPW) and targeted minimum loss–based estimation (TMLE). Standard errors and CIs were not computed for the G-computation estimator, since this would require many bootstrap iterations across simulation replicates.

### Inverse probability weighting (IPW)

The IPW estimator for the ATE ($\psi$) uses predictions $\hat{\pi}(x)$ from the treatment nuisance model (propensity score). IPW estimators can be viewed as taking the difference in weighted outcome means between treatment groups, with weights computed as the inverse probability of treatment conditional on covariates.

We estimated treatment effects from the coefficient on $A$ in a weighted least-squares regression using stabilized weights (w), which is algebraically equivalent to the Hájek IPW estimator.[61] Hájek normalizes the Horvitz-Thompson form[62] by dividing each unit's weight by the sum of all weights in that treatment group.

$$\widehat{w}_i = A_\mathrm{i}\frac{\widehat{Pr}(A=1)}{\hat{\pi}(X_\mathrm{i})} + (1-A_\mathrm{i})\frac{1-\widehat{Pr}(A=1)}{1-\hat{\pi}(X_\mathrm{i})} = \begin{cases} \dfrac{\widehat{Pr}(A=1)}{\hat{\pi}(X_\mathrm{i})} \; for\ A=1 \\ \dfrac{1-Pr(A=1)}{1-\hat{\pi}(X_\mathrm{i})} \; for\ A=0 \end{cases}$$

*Hájek IPW estimator for the ATE:*

$$\hat{\psi}_{IPW_{Hajek}} = \frac{\sum_{i=1}^{n}\widehat{w}_\mathrm{i}A_iY_i}{\sum_{i=1}^{n}\widehat{w}_\mathrm{i}A_i} - \frac{\sum_{i=1}^{n}\widehat{w}_\mathrm{i}(1-A_i)\,Y_i}{\sum_{i=1}^{n}\widehat{w}_\mathrm{i}(1-A_i)} = \frac{\sum_{i=1}^{n}A_i\,Y_i/\hat{\pi}(X_i)}{\sum_{i=1}^{n}A_i/\hat{\pi}(X_i)} - \frac{\sum_{i=1}^{n}(1-A_i)\,Y_i/\{1-\hat{\pi}(X_i)\}}{\sum_{i=1}^{n}(1-A_i)/\{1-\hat{\pi}(X_i)\}}$$

### G-computation

The ATE can also be estimated via the G-computation estimator, using predictions from the outcome model nuisance fits, $\hat{\mu}_1(x)$ and $\hat{\mu}_0(x)$.

$$\hat{\psi}_{G-comp} = \frac{1}{n}\sum_{i=1}^{n}\hat{\mu}_1(X_i) - \frac{1}{n}\sum_{i=1}^{n}\hat{\mu}_0(X_i)$$

### Augmented IPW (AIPW)

Doubly robust estimators such as the Augmented IPW (AIPW) estimator use predictions from both the treatment (propensity score) and outcome nuisance fits [20,19,15].

*For the ATE:*

$$\hat{\psi}_{AIPW} = \frac{1}{n}\sum_{i=1}^{n}\left[\hat{\mu}_1(X_i) + \frac{A_i}{\hat{\pi}(X_i)}\{Y_i - \hat{\mu}_1(X_i)\}\right] - \frac{1}{n}\sum_{i=1}^{n}\left[\hat{\mu}_0(X_i) + \frac{1-A_i}{1-\hat{\pi}(X_i)}\{Y_i - \hat{\mu}_0(X_i)\}\right]$$

### Targeted minimum loss–based estimation (TMLE)

Targeted minimum loss–based estimation (TMLE) (also known as "targeted maximum likelihood estimation") is another doubly robust estimator using both treatment model and outcome model nuisance fits. TMLE is asymptotically equivalent to AIPW under suitable regularity conditions but can have improved finite-sample performance.[26,27] TMLE has a separate targeting step during which predictions from the outcome models are "updated" to optimize the bias-variance tradeoff for the target estimand (specifically, to solve the efficient influence function estimating equation).[21,27]

$$\hat{\psi}_{TMLE} = \frac{1}{n}\sum_{i=1}^{n} \hat{\mu}_1^*(X_i) - \frac{1}{n}\sum_{i=1}^{n} \hat{\mu}_0^*(X_i)$$

where $\hat{\mu}_1^*(X_i)$ and $\hat{\mu}_0^*(X_i)$ are the updated (or corrected) outcome predictions. Note that continuous outcomes must be rescaled between [0,1] and their predictions bounded before the logistic targeting step, and then transformed back to the original outcome scale; this happens internally within our implementation using the *tmle* R package (v2.1.1).[40,63] Our notation here assumes both the outcome $Y$ and predictions are rescaled values. To obtain updated outcome predictions, we first fit a no-intercept logistic regression model for the rescaled outcome using the logit of the initial predicted outcomes under each individual's observed treatment as an offset term.

$$logit(E[Y_i|A_i, X_i]) = logit\left(\hat{\mu}_{A_i}(X_i)\right) + \epsilon_1 \hat{H}_1(A_i, X_i) + \epsilon_0 \hat{H}_0(A_i, X_i)$$

where $\hat{\mu}_{A_i}(X_i) = A_i \hat{\mu}_1(X_i) + (1 - A_i)\hat{\mu}_0(X_i)$ and $\hat{H}_1, \hat{H}_0$ are the "clever covariates":

$$\hat{H}_1(A_i, X_i) = \frac{A_i}{\hat{\pi}(X_i)}$$

$$\hat{H}_0(A_i, X_i) = \frac{1 - A_i}{1 - \hat{\pi}(X_i)}$$

Using the estimated $\hat{\epsilon}_1$ $and$ $\hat{\epsilon}_0$, the initial $\hat{\mu}_1(X_i)$ and $\hat{\mu}_0(X_i)$ estimates are then "updated", for every individual under both treatment levels.

$$\hat{\mu}_1^*(X_i) = expit\{logit(\hat{\mu}_1(X_i)) + \hat{\epsilon}_1/\hat{\pi}(X_i)\}$$

$$\hat{\mu}_0^*(X_i) = expit\{logit(\hat{\mu}_0(X_i)) + \hat{\epsilon}_0/(1 - \hat{\pi}(X_i))\}$$

And the $\hat{\psi}_{TMLE}$ ATE estimate is computed as the difference in updated conditional outcome means, $\frac{1}{n}\sum \hat{\mu}_1^*(X_i) - \frac{1}{n}\sum \hat{\mu}_0^*(X_i)$, after transforming predictions back to the original scale. The variance was estimated from the efficient influence curve.

### A3. Modeling nuisance functions

We explored a range of learning algorithms for estimating the $\pi(x)$ and $\mu_a(x)$ models, spanning parametric regression, standalone ML learners, pre-trained tabular foundation models, and richly-specified Super Learner ensembles. We truncated estimated propensity scores below 0.025 or above 0.975 for all nuisance learners and estimators, including the oracle nuisances. All learners except for parametric regression fit separate conditional outcome models among the treated, $\mu_1(x) = E[Y \mid A = 1, X = x]$, and untreated $\mu_0(x) = E[Y \mid A = 0, X = x]$. For all learners undergoing hyperparameter tuning, exposure and outcome nuisance models were tuned separately. Cross-validation for internal tuning used the same fold counts ($K = 10{,}5{,}5$ $for$ $n = 200, 1200, 5000$). Learner details are provided in **Table S1**.

**Table S1.** Nuisance learner algorithm details

| Learner | Details |
|---|---|
| Parametric † | Generalized Linear Models (GLM) via glm package; Logistic for treatment model and linear for outcome model. This standard parametric learner will be correctly specified by-definition in the simple confounding scenario of simulation 1. For ATE and ATT estimation, the parametric learner fit one pooled outcome model across both treatment groups. |
| Random Forest † | Via ranger package; num.trees=500, mtry= $floor(\sqrt{p})$, min.node ∈ {30, 60}, where p is the number of covariates. For simulation 1, mtry=2, which corresponds to the hyperparameter configuration from the SL Flexible Random Forest base learner. For simulation 2, mtry=8. |
| Highly Adaptive Lasso (HAL) † | Via hal9001 package; max_degree=2, smoothness_orders∈ $(0,1)$, reduce_basis=NULL. Selected as a modern learner considered state-of-the-art for semi-parametric causal inference. Due to prohibitive run time at the largest simulation 1 sample size, the HAL variant used a fixed smoothness_order=1 with no smoothness selection (the lasso penalty was still chosen by 5-fold cross-validation). The simulation 2 variant used smoothness_order=1, as well as num_knots = c(25,10) and corRank pre-screening to the top 10 covariates prior to fitting HAL nuisance model. |
| TabPFN † | TabPFN (version v3 weights) accessed via Prior Lab's cloud API (tabpfn-client 0.3.0) and, separately, run locally on CPU (tabpfn 8.0.8 with Python torch 2.11.0) for CPU run time measurements. TabPFN was used out-of-the-box with default settings in both cases (n_estimators = 8, random_state=0, no post-hoc calibration, no fine tuning) |
| SL Flexible † | SL ensemble with base learners (each considering main terms and all 2-way interactions):<br>• Random forest – via SL.ranger. num.trees=500, mtry=2, min.node ∈ $\{30, 60\}$<br>• XGBoost – via SL.xgboost. num.trees=500, max.depth=4, learning.rate=0.1, min.node ∈ {30, 60}<br>• Generalized Additive Model (GAM) – via SL.gam. deg.gam ∈ {3:8}, maxit = 50, bf.maxit = 50<br>SL Flexible was the top performing model in the simulation 1 setting as described by Naimi et al., used here as a competitive baseline. For simulation 1, each base learner used main features and all 2-way interactions as inputs. The higher-dimensional simulation 2 used only main terms as inputs, $mtry = floor(\sqrt{80}) = 8$ for the random forest base learners, and a corRank top-20 screening algorithm for each nuisance fit. |
| SL Lightweight † | SL ensemble with base learners:<br>• Parametric Generalized Linear Model -- via SL.GLM. Logistic for treatment model, linear for outcome model.<br>• Stepwise regression with pairwise interactions -- via SL.step.interaction; defaults (scope= Y~.^2, direction = "both", k=2, steps=1000).<br>• Multivariate adaptive regression splines (MARS) -- via SL.earth; package defaults.<br>• Mean -- via SL.mean<br>SL Lightweight was among the best performing learners in simulation 1 setting from a previous study by Balzer et al., used here as a competitive baseline, representing an ensemble that contains the true simulation 1 DGP. In Simulation 2, the stepwise regression with pairwise interactions base learner used a corRank top-10 screening algorithm to accommodate the higher-dimensional setting. |
| SL Default † | SL ensemble with base learners:<br>• Parametric Generalized Linear Model (GLM) – via SL.glm. Logistic for treatment model, linear for outcome model.<br>• Bayesian Additive Regression Trees (BART) -- smoothing parameter k= 2 for models via tmle.SL.dbarts2, k = 0.5 for model via tmle.SL.dbarts.k.5; otherwise defaults.<br>• Generalized Additive Model (GAM) -- via SL.gam. Used for treatment model only. deg.gam=2, maxit = 50, bf.maxit = 50.<br>• LASSO -- via SL.glmnet. Used for outcome models only. Lambda chosen via cross-validation (nlambda = 100).<br>The "diverse" ensemble default in the tmle and SuperLearner R packages. For the higher dimensional simulation 2 setting, corRank top-20 screening algorithms were used for the BART and GAM base learners. |
| XGBoost* | Via xgboost package; nrounds=500, max_depth=4, eta=0.1, min_child_weight∈{30, 60}. |
| GAM* | Via gam through SL.gam wrapper; deg.gam=4, maxit = 50, bf.maxit = 50 |
| BART* | Via dbarts; k (smoothing parameter)=2, sigdf=3, sigquant=0.9, power=2, base=0.95, ntree=200, ndpost=500, nskip=100, numcut=100. |
| LASSO* | Via glmnet::cv.glmnet; pre-processing of predictors to consider main effects plus all 2-way interactions. Params: alpha=1, lambda=lambda.min via internal cross-validation, nlambda=100, maxit=100000. |
| KNN* | Via R package caret::knnreg $k \in \{5,10,15,20,30,50\}$, via internal cross-validation. |
| SL Lightweight + TabPFN* | SL ensemble with base learners: TabPFN v3 + SL Lightweight base learner library. Explores impact of adding tabPFN to lightweight ensemble with parametric and other less adaptive learners. |
| SL Parametric + TabPFN* | SL ensemble with base learners: TabPFN v3 + Parametric regressions (GLMs). Explores impact of adding tabPFN to ensemble with parametric regression, specifically to understand how this influences finite-sample performance within the simple (parametric) scenario of simulation 1. |
| Mitra (PFN)* | Mitra is an alternative PFN-based tabular foundation model, with a different approach for enhancing the synthetic priors included during pre-training.[38] We used the 2nd-generation Mitra weights[64] (Mitra-v2; mitra-classifier-2 and mitra-regressor-2), implemented via AutoGluon (autogluon.tabular 1.6.2). Mitra parameters were modified slightly from defaults to enable fair comparison with TabPFN: n_estimators = 8, fine_tune = FALSE; default settings otherwise. |
| TabPFN, streamlined (single pass)* | TabPFN-v3's default configuration internally aggregates multiple predictions for stability (default: n_estimators=8), although this does not include other model classes and involves no weight updates/tuning; it simply averages predictions across 8 forward passes with the same frozen model, each on a randomly permuted and preprocessed copy of the data (e.g., re-shuffling column and/or row order). To investigate how this affects performance, we implemented a TabPFN variant using only a single forward pass (n_estimators=1). |

*BART = Bayesian Additive Regression Trees, XGBoost = Extreme Gradient Boosting, GAM = Generalized Additive Model, KNN = K-Nearest Neighbors, LASSO = Least Absolute Shrinkage and Selection Operator, PFN = prior-data fitted network, SL = super learner.*

† Primary learners, selected a priori, emphasized in the main manuscript and implemented across both simulation settings and all sample sizes.

*Secondary learners; only implemented for simulation 1.

**Note.** All SL ensembles were implemented as convex ensembles with the R SuperLearner package, with method="method.NNLS", using gaussian outcome models fit separately for each treatment arm and binomial treatment models. method.NNLS chooses ensemble weights via non-negative least squares on the cross-validated predictions, then normalizes to sum to one. For learners using screening algorithms in the higher-dimensional simulation 2, screening was applied separately within each nuisance fit and cross-fitting fold. All nuisance learners converged successfully across all runs.

## A4. Cross-fitting

Cross-fitting, a version of sample-splitting, is an approach for controlling own-observation overfitting bias by decoupling nuisance fitting from causal estimation. This helps to side-step Donsker-type complexity conditions and permits a wider range of data-adaptive algorithms as nuisance learners without comprimising statistical inference, provided that convergence requirements and other regularity conditions are satisfied.[15,21] Cross-fitting has favorable theoretical properties and has demonstrated empirical benefits in simulations.[21] However, it is more computationally intensive and, in finite samples, can increase variance and instability because each model is fit on a reduced subset of the data. These drawbacks may be amplified when nonparametric learners - and especially ensemble methods - use cross-validation for tuning and model selection, as this introduces an additional layer of nested cross-validation within each sample split. Because TabPFN's in-context learning does not require dataset-specific tuning, the computational and finite-sample trade-offs of cross-fitting may also differ from those observed with conventional learners.

To investigate this, we estimated target quantities both with and without cross-fitting of the underlying nuisance models. Here, we use the "Double/debiased DML-2" pooled cross-fitting approach from Chernozhukov et al.[15] For the cross-fitting variants, the data were randomly partitioned into $K$ non-overlapping folds ($K = 10{,}5{,}5$ for $n = 200{,}1000/1200{,}5000$). For each fold, nuisance models for treatment and outcome were trained on the remaining $K - 1$ folds and used to generate out-of-sample predictions on the held-out fold. Most learners underwent further levels of inner cross-validation for model tuning and super learner ensembling within each cross-fitting partition. This procedure was repeated across all folds so that every observation received nuisance predictions from models trained on data that excluded that observation. Finally, nuisance predictions were pooled across folds and then plugged into the estimators to obtain estimates for the target causal quantity. Note that this implementation is related to, but distinct from, cross-validated TMLE (CV-TMLE). To reduce learner-specific sensitivity to particular sample splits, we used the same deterministic fold assignments so that learners saw identical fold partitions for a given simulated dataset. (Cross-fit TabPFN was an exception, due to differences in fold assignment within the python API implementation).

## A5. Simulation 1 Details: Fully synthetic data with simple and complex confounding

In our first fully synthetic simulation, we used the simple (parametric) and complex (non-parametric) confounding designs described by Naimi et al.,[16] originally adapted from Kang and Schafer.[33] These simulation settings were used to explore the tradeoffs and practical challenges associated with using flexible nuisance learners for causal estimation under different confounding structures. For both the simple and complex confounding scenarios, outcome and treatment variables were generated from parametric functions of four independent standard normal confounding variables ($X_1, X_2, X_3, and\ X_4$). The treatment variable was generated with a logistic model:

$$P(A = 1|X_1, X_2, X_3, X_4) = expit\{-1 + ln(1.75)X_1 + ln(1.75)X_2 + ln(1.75)X_3 + ln(1.75)X_4\}$$

And the continuous outcome variable with a linear model:

$$Y = 120 + 6A + 3X_1 + 3X_2 + 3X_3 + 3X_4 + \epsilon$$

With i.i.d errors, $\epsilon \sim N(0{,}36)$. The true average treatment effect (ATE) was $\psi = 6$ in both scenarios, with no heterogeneity. Using this structure, we generated 200 Monte Carlo replicates for each of three sample sizes ($n = 200{,}1200{,}5000$). To ensure all nuisance learners and estimators were evaluated on equal footing, we first generated and saved all $K = 200$ simulated datasets within each sample size and scenario, and then implemented all nuisance learners and estimators across the same datasets. (Exceptions include the TabPFN nuisance fits at $n = 5000$, which used an independently generated replicate set from the same underlying DGP, therefore maintaining comparability at the aggregate level).

In the simple confounding scenario, models were fit directly on the observed $A$, $Y$, and $X$ values from each simulation run. In the complex confounding scenario, only the transformed confounders $Z1 - Z4$ were available for model development, inducing complex non-linear confounding.

$$Z_1 = exp(X_1/2)$$
$$Z_2 = X_2/(1 + exp(X_1)) + 10$$
$$Z_3 = \left(\left(\frac{X_1 X_3}{25}\right) + 0.6\right)^3$$
$$Z_4 = (X_2 + X_4 + 20)^2$$

Simulation 1 targeted only the ATE estimand. Note that we were unable to precisely reproduce the results reported in Naimi et al.[16] and suspect that their underlying simulation used a slightly different specification (specifically, a multiplicative rather than additive transformation of $Z_4$). This discrepancy does not appear to affect the conclusions of Naimi et al. nor the original motivation for choosing this simulation design. Our simulation setup also aligns with the original $Z_4$ transformation from Kang and Schafer.[33]

**Evaluation metrics for simulation 1**

The primary evaluation metrics for the ATE included bias, root mean squared error (RMSE), 95% confidence interval (CI) coverage, and run time. Bias was computed as the absolute value of the average per-simulation difference between the estimated and ground-truth causal effect, expressed as a percentage of the true causal effect: $\frac{\left|\frac{1}{K}\sum_{k=1}^{K}(\hat{\psi}_k - \psi)\right|}{\psi}$ x $100$, where $\hat{\psi}_k$ is the ATE estimate from simulation $k = \{1, \ldots, K\}$ for $K = 200$ simulation replicates, and $\psi$ is the true ATE. We report RMSE in absolute terms and as a percentage of the Oracle RMSE, which represents the error that remains when nuisance estimation is perfect (i.e., when estimators use the true nuisance functions). Coverage was defined as the proportion of Wald-type confidence intervals that included the true causal effect. In figures and tables, we denote coverage cells that are not significantly below the 0.95 nominal level via a one-sided z-test at $\alpha = 0.05$ using the Monte Carlo standard error under the null: $\text{coverage} \geq 0.95 - 1.645\sqrt{(0.95\ (0.05/\text{K}))}$. At $K = 200$ simulations, this means that coverage values ≥ $0.9246$ are (informally) within Monte Carlo simulation noise of nominal. Run times were measured as the average wall time to fit both the propensity score and outcome mean(s) nuisance fits, across 35 sims of the complex scenario ($n = 1200$). CPU times used a local laptop with 12 CPU cores and 24 GB memory; GPU times used an A100 GPU.

As secondary metrics, we report the average confidence interval width and the SE calibration ratio, defined as the average model-reported SE (from the efficient influence function or robust variance estimator) divided by the empirical SD of the point estimate ($\hat{\psi}$) across $K = 200$ simulations: 1 indicates perfect calibration, <1 over-confident, >1 conservative.

$$\text{SE ratio} = \frac{\frac{1}{K}\sum_{k=1}^{K}\widehat{SE}_k}{\sqrt{\frac{1}{K-1}\sum_{k=1}^{K}\left(\hat{\psi}_k - \frac{1}{K}\sum_{k=1}^{K}\hat{\psi}_k\right)^2}}$$

We also evaluated baseline covariate balance with the ATE IPW estimator as an additional proxy for confounding control performance, within the complex scenario of simulation 1 at $n = 1200$. Absolute standardized mean differences (ASMD)[65] between treated $(A = 1)$ and untreated $(A = 0)$ observations were computed for each covariate $j$ as:

$$\text{ASMD}_j = \frac{\left|\frac{\sum_{i=1}^{n}\hat{w}_i A_i x_{i,j}}{\sum_{i=1}^{n}\hat{w}_i A_i} - \frac{\sum_{i=1}^{n}\hat{w}_i(1-A_i)x_{i,j}}{\sum_{i=1}^{n}\hat{w}_i(1-A_i)}\right|}{\sqrt{(\hat{\sigma}_{j,1}^2 + \hat{\sigma}_{j,0}^2)/2}}$$

where $\hat{w}_i$ is the estimated stabilized inverse probability weight for individual $i$, $x_{i,j}$ is the value of covariate j for individual $i$, and $\hat{\sigma}^2{}_{j,a}$ denotes the unweighted estimated sample variance of covariate $j$ in treatment group $A = a$, for one Monte Carlo simulation draw. For each nuisance learner, ASMDs were averaged over the four covariates within each simulation and then over $50$ simulation runs (a subset of the full $K = 200$). Within the complex scenario, we computed ASMDs for the "true" latent confounders $\{X_1, X_2, X_3, X_4\}$ and the observed transformed confounders provided to each nuisance learner $\{Z_1, Z_2, Z_3, Z_4\}$. Covariate balance for the true generating covariates is the relevant proxy for whether IPW adequately controlled for confounding, whereas balance for the transformed confounders represents the naïve diagnostic an analyst would see without considering the true functional forms.

### A6. Simulation 2 Details: Plasmode simulations across diverse data generating processes

Simulation 2 evaluated TabPFN and diverse ensembles across many DGPs to understand if conclusions are robust across different hypothetical data structures and causal estimands we may encounter in practice. The second simulation set-up used data generating processes (DGPs) from the Atlantic Causal Inference Conference (ACIC) 2016 data analysis competition.[42] Unlike simulation 1, which was fully synthetic, the ACIC plasmode scenario used real covariate data drawn from the Collaborative Perinatal Project, a longitudinal study of pregnancy outcomes.[41] The covariate matrix included 58 covariates of mixed type (23 continuous, 27 count, 5 binary, and 3 categorical), contributing 80 total degrees of freedom. From this fixed covariate matrix, we generated 44 distinct DGP settings by using different combinations of "knobs" to simulate treatment and outcome processes, defined by the ACIC organizers and available through the R package *aciccomp2016*. Knobs included the degree of nonlinearity for treatment assignment (linear, polynomial, step functions) and outcome models (linear, exponential, step functions); the proportion treated (0.35 or 0.65); alignment between treatment and outcome predictors to determine the magnitude of confounding (none, medium, high); the degree of treatment effect heterogeneity (none, medium, high); and propensity score overlap (full or partial overlap). The 44 distinct DGP settings are shown in **Table S2**. Note we include several de-novo knob combinations not explored in the original ACIC competition, selected to add additional ATE-identified settings.

For each simulation 2 estimand, we limited to DGPs for which the respective estimand was identifiable. Specifically, the ATE estimand was evaluated only in the 35 DGP settings with full propensity score overlap. Because the partial overlap knob introduced violations to the ATE positivity condition but not the ATT partial positivity condition, the ATT estimand was evaluated across all 44 DGPs.

For each DGP setting - defined with a fixed knob configuration – we ran $K = 50$ Monte Carlo simulation replicates. The per-replicate process was as follows:

1. Randomly draw treatment and outcome surfaces ($\pi\ and\ \mu$) from the fixed covariate matrix (4802 x 58), with functional forms and calibration determined by the setting's knobs.
2. Generate treatment $A_i$ and outcomes $Y_i$ values for all $N = 4802$ individuals via random draws from the fixed surfaces in (1).
3. For computational tractability, further sub-sample without replacement to obtain an n = 1,000 analytic sub-sample for nuisance estimation and effect estimation.

The target estimands for simulation 2 were the sub-sample ATE and ATT. Within each fixed DGP setting, we denote the true conditional average treatment effect (CATE) function for replicate $k$ as $\tau_k(x) = \mu_{1,k}(x) - \mu_{0,k}(x)$. The true sub-sample ATE and ATT for each replicate, which served as the grading targets, were computed by averaging over the CATE evaluated at each unit's observed covariates $\tau_k(X_i)$, within each sub-sample. Letting $\mathcal{S}_k$ denote the $n = 1000$ sub-sample drawn from the full $N = 4802$ cohort for each Monte-Carlo draw $k$, the sub-sample ATE is:

$$\psi^{sub}{}_k = \frac{1}{n}\sum_{\mathrm{i}\in\mathcal{S}_k} \mathrm{\tau_k}(X_i)$$

and the corresponding ATT is:

$$\delta_k^{sub} = \frac{1}{n_{1k}}\sum_{i\in\mathcal{S}_k} A_{i.k}\,\tau_k(X_i), \qquad where\ n_{1k} = \sum_{i\in\mathcal{S}_k} A_{i.k}$$

Across all identified DGPs, the mean true sub-sample ATE ranged from 3.3 to 4.2 and the mean true sub-sample ATT from 3.3 to 4.1. Like the original ACIC competition described by Dorie et al.,[42] we targeted sample rather than population estimands. However our approach differed in three ways: (i) We analyzed the n=1,000 sub-sample (step 3) instead of the full $N = 4802$ cohort, (ii) we added an ATE target estimand to the original ATT-only estimand, and (iii) we computed targets in each replicate as the in-sample average of the true conditional mean effects $\mathrm{\tau_k}(X_i)$ instead of the realized individual effects.[66]

*Evaluation for simulation 2*

Bias and RMSE were graded against their replicate-specific sub-sample targets to remove variation from the random sub-sampling step. Letting $\hat{\theta}_k$ denote the estimate from sub-sample $k$ ($\hat{\psi}_\mathrm{k}$ or $\hat{\delta}_k$) and $\theta_k^{\mathrm{sub}}$ the corresponding sub-sample grading target, the bias and RMSE metrics for each fixed DGP configuration across $K = 50$ replicates were computed as follows:

$$|bias| = \left|\frac{1}{K}\sum_{k=1}^{K}\left(\hat{\theta}_k - \theta_k^{sub}\right)\right|$$

$$\text{Relative bias as a percentage of target} = 100 \text{ x} \frac{|bias|}{\left|\frac{1}{K}\sum_{k=1}^{K}\theta_k^{sub}\right|}$$

$$RMSE = \sqrt{\frac{1}{K}\sum_{k=1}^{K}\left(\hat{\theta}_k - \theta_k^{sub}\right)^2}$$

Macro-averages were calculated as unweighted averages across DGPs.

Confidence interval coverage and SE calibration ratios were used as secondary evaluation metrics for simulation 2, given we used only 50 replicates per DGP and due to complications introduced from the multi-level sampling procedure. Specifically, the influence function-based variance estimators used for AIPW and TMLE target population parameters, and may therefore over-estimate the sample estimand's variance, particularly under heterogeneity.[67] We therefore graded coverage and SE calibration against full-cohort targets, $\theta_k^{cohort}$, computed as the average of $\tau_k(X_i)$ over all units (or all treated units) in the full $N = 4802$ covariate matrix before sub-sampling. For a given DGP setting, coverage was defined as the proportion of the $K = 50$ replicate-level 95% confidence intervals, $\hat{\theta}_k \pm 1.96 \text{ x } \widehat{\text{SE}}_k$, that contained $\theta_k^{cohort}$ from the same replicate, where $\widehat{\text{SE}}_k$ is the model-reported (influence-function-based) standard error from replicate $k$. We evaluated whether intervals contained true targets separately within each replicate, since the treatment and outcome surfaces were re-drawn even within a fixed DGP configuration. Similarly, SE calibration ratios for each DGP were calculated as the mean model-reported SE across $K = 50$ replicates, divided by the empirical standard deviation of the estimation errors relative to each replicate's full-cohort target:

$$\text{SE calibration ratio} = \frac{\frac{1}{K}\sum_{k=1}^{K}\widehat{SE}_{\text{k}}}{\sqrt{\frac{1}{K-1}\sum_{k=1}^{K}\left(b_k - \bar{b}\right)^2}}, \qquad where\ \ b_k = \hat{\theta}_{\text{k}} - \theta_k^{\text{cohort}}$$

These metrics will be mildly conservative since the $N = 4802$ cohort is itself a finite proxy for the so-called superpopulation, and because the model-reported SE does not account for the random sub-sampling without replacement. For instance, for the oracle TMLE estimation of the ATE across the 35 full-overlap DGPs, coverage was 96.1% and the SE calibration ratio was 1.08.

*Nuisance learners for simulation 2*

For Simulation 2, we restricted attention to the primary standalone learners and SL ensemble algorithms. Several modifications to learner configurations were required to accommodate the higher-dimensional setting ($p = 80$). SL Flexible was limited to main effects only, and both SL Lightweight and SL Default incorporated screening algorithms to select a subset of predictors before fitting p-sensitive base learners (screening was used for stepwise regression with pairwise interactions in SL Lightweight, and for the BART and GAM base learners in SL Default). See Table S1 for further detail.

**Table S2**. Data-generating process (DGP) settings evaluated in plasmode simulation 2

| DGP ID | ACIC ID | Tx model | Outcome model | Alignment | Tx prev | Overlap | Heterogeneity |
|---|---|---|---|---|---|---|---|
| 1* | - | linear | linear | 0.75 | 0.35 | full | high |
| 2* | - | linear | exponential | 0.75 | 0.35 | full | high |
| 3* | - | polynomial | linear | 0.75 | 0.35 | full | high |
| 4* | - | linear | exponential | 0.75 | 0.65 | full | high |
| 5* | - | linear | step | 0.75 | 0.35 | full | high |
| 6* | - | step | linear | 0.75 | 0.35 | full | high |
| 7* | - | linear | step | 0.25 | 0.35 | full | med |
| 8* | - | step | linear | 0.25 | 0.35 | full | med |
| 9 | 4 | polynomial | exponential | 0.75 | 0.35 | full | high |
| 10 | 28 | polynomial | step | 0.25 | 0.35 | full | high |
| 11 | 29 | polynomial | step | 0.75 | 0.35 | full | med |
| 12 | 30 | polynomial | step | 0.75 | 0.35 | full | high |
| 13 | 32 | polynomial | exponential | 0.25 | 0.35 | full | high |
| 14 | 33 | polynomial | exponential | 0.75 | 0.35 | full | med |
| 15 | 41 | polynomial | step | 0.25 | 0.65 | full | high |
| 16 | 42 | polynomial | step | 0.75 | 0.65 | full | med |
| 17 | 43 | polynomial | step | 0.75 | 0.65 | full | high |
| 18 | 45 | polynomial | exponential | 0.25 | 0.65 | full | high |
| 19 | 46 | polynomial | exponential | 0.75 | 0.65 | full | med |
| 20 | 47 | polynomial | exponential | 0.75 | 0.65 | full | high |
| 21 | 56 | step | step | 0.25 | 0.35 | full | high |
| 22 | 57 | step | step | 0.75 | 0.35 | full | med |
| 23 | 58 | step | step | 0.75 | 0.35 | full | high |
| 24 | 60 | step | exponential | 0.25 | 0.35 | full | high |
| 25 | 61 | step | exponential | 0.75 | 0.35 | full | med |
| 26 | 62 | step | exponential | 0.75 | 0.35 | full | high |
| 27 | 71 | step | step | 0.25 | 0.65 | full | high |
| 28 | 72 | step | step | 0.75 | 0.65 | full | med |
| 29 | 73 | step | step | 0.75 | 0.65 | full | high |
| 30 | 75 | step | exponential | 0.25 | 0.65 | full | high |
| 31 | 76 | step | exponential | 0.75 | 0.65 | full | med |
| 32 | 77 | step | exponential | 0.75 | 0.65 | full | high |
| 33* | - | linear | linear | 0.75 | 0.35 | full | none |
| 34* | - | polynomial | exponential | 0.75 | 0.35 | full | none |
| 35* | - | step | exponential | 0.75 | 0.35 | full | none |
| 36 | 2 | polynomial | exponential | 0.75 | 0.35 | one-term | none |
| 37 | 3 | linear | linear | 0.75 | 0.35 | one-term | none |
| 38 | 5 | linear | exponential | 0.75 | 0.35 | one-term | high |
| 39 | 6 | polynomial | linear | 0.75 | 0.35 | one-term | high |
| 40 | 7 | polynomial | exponential | 0.75 | 0.35 | one-term | high |
| 41 | 8 | polynomial | exponential | 0 | 0.35 | one-term | high |
| 42 | 9 | step | step | 0.75 | 0.35 | one-term | high |
| 43 | 24 | polynomial | step | 0.75 | 0.35 | one-term | high |
| 44 | 54 | step | exponential | 0.75 | 0.35 | one-term | high |

* = Indicates custom DGP setting not included in the original 2016 Atlantic Causal Inference Competition (ACIC). Custom DGPs were created using the same ACIC machinery to ensure adequate representation of simpler (linear) structures in the full overlap scenarios.

Data generating process (DGP) ID: Corresponds to DGP identifiers used in manuscript figures.

ACIC ID: Corresponds to label used in original 2016 ACIC competition.[42]

True Tx model = functional forms for governing A-X relationships; True outcome model = functional forms for governing A-X-Y relationships; Treatment prev = exposed proportion; Alignment = Proportion of covariates related to both A and Y, to determine magnitude of confounding; Heterogeneity = none, medium, or high.

Overlap (in PS) = Full vs partial overlap. Partial overlap settings labeled "one-term" overlap in ACIC competition; partial non-overlap created covariate regions with untreated but no treated individuals, so the ATT remains identified.

### A7. Software

Analyses were executed in R (version 4.5.2) and Python (version 3.11), and study code is available on GitHub: https://github.com/a-weckstein/Tabular-FM-causal-simulations, https://github.com/a-weckstein/ACIC-plasmode-pipeline. TabPFN-v3 nuisance fits were executed via Prior Labs' hosted API, which allows users to submit tabular prediction queries with a limited quota of free predictions per day. The API and model weights were made available for non-commercial research use under Prior Labs' open license. We also executed select runs locally and via Google Colab, for GPU timings. Authors used Claude (Opus v5-v5.5, Fable v5) for assistance with code review, to double program core simulation results, and for proofreading and making minor language edits to the manuscript text. The initial manuscript was drafted entirely by coauthors, and authors have reviewed and take responsibility for all content in the final manuscript.

# Supplement 2. Extended Results

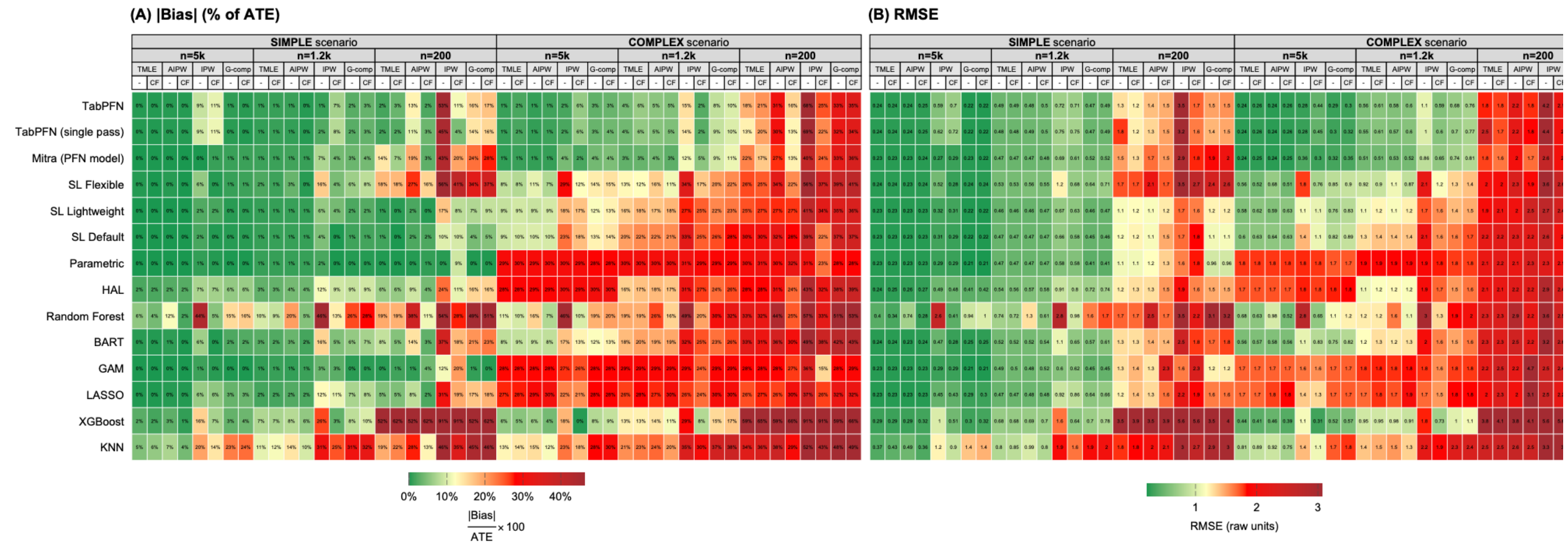

**Figure S1. Bias and RMSE for estimation of the average treatment effect in synthetic simulation 1 for all estimators**. Each cell was computed from 200 simulation replicates within each configuration (estimator × scenario × sample size × cross-fitting), displaying each nuisance learner's (A) absolute bias as percentage of true ATE and (B) RMSE, without oracle scaling. Color scales are capped at the 95th-percentile value for each metric. *Abbreviations: AIPW = augmented inverse probability weighting,* ATE = *average treatment effect*, *CF = cross-fit, HAL = highly-adaptive LASSO, RMSE = root mean squared error, SL = super learner ensemble, TMLE = targeted minimum loss-based estimation*

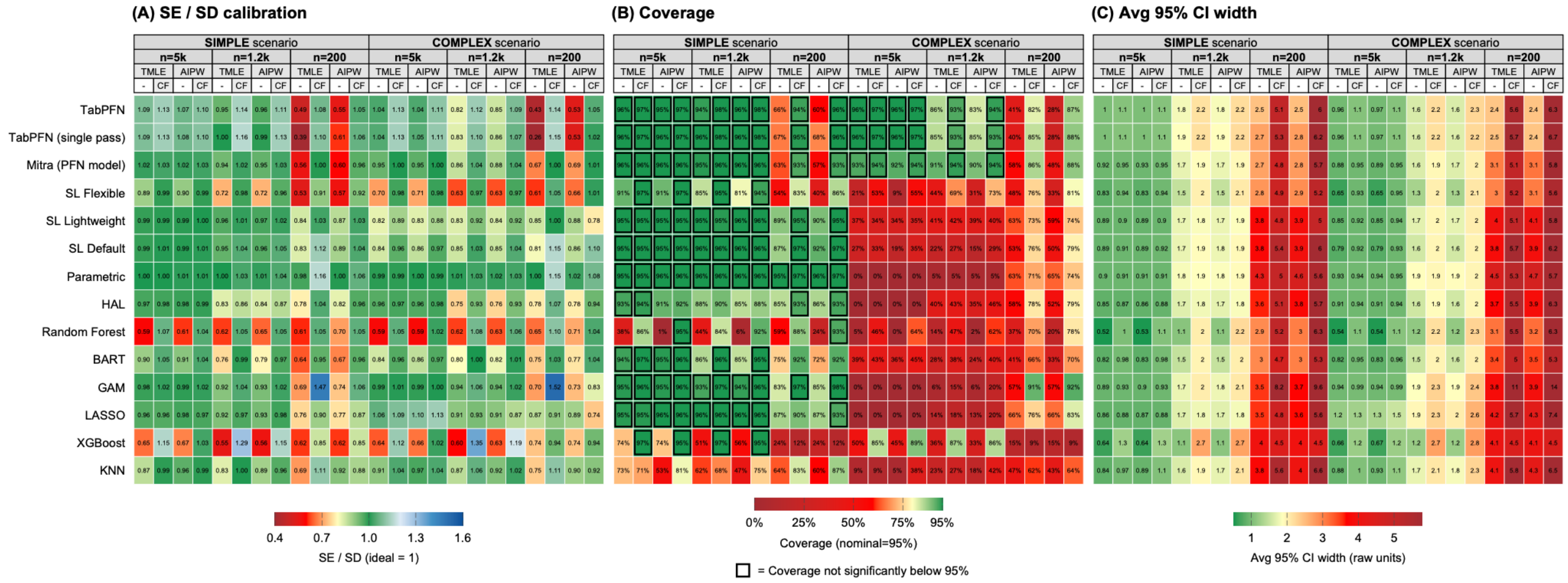

**Figure S2. SE calibration, coverage, and confidence interval width for doubly robust estimation of the average treatment effect in synthetic simulation 1.** Each cell was computed from 200 simulation replicates within each configuration (estimator × scenario × sample size × cross-fitting), displaying each nuisance learner's (A) SE calibration ratio (1 indicates perfect calibration, <1 over-confident, >1 conservative), (B) 95% confidence interval coverage, and (C) average width of the 95% confidence interval. Read down rows to compare learners and across columns to compare estimators and configurations. Color scales are capped at the 95th-percentile value for each metric. Coverage defined as the proportion of 200 replicate-level 95% confidence intervals containing the true ATE. Bordered cells indicate coverage not significantly below nominal (95%), via a one-sided z-test using the Monte Carlo standard error under the null. *Abbreviations: AIPW = augmented inverse probability weighting,* ATE = *average treatment effect*, *CF = cross-fit, HAL = highly-adaptive LASSO, SD = standard deviation, SE = standard error, SL = super learner ensemble, TMLE = targeted minimum loss-based estimation.*

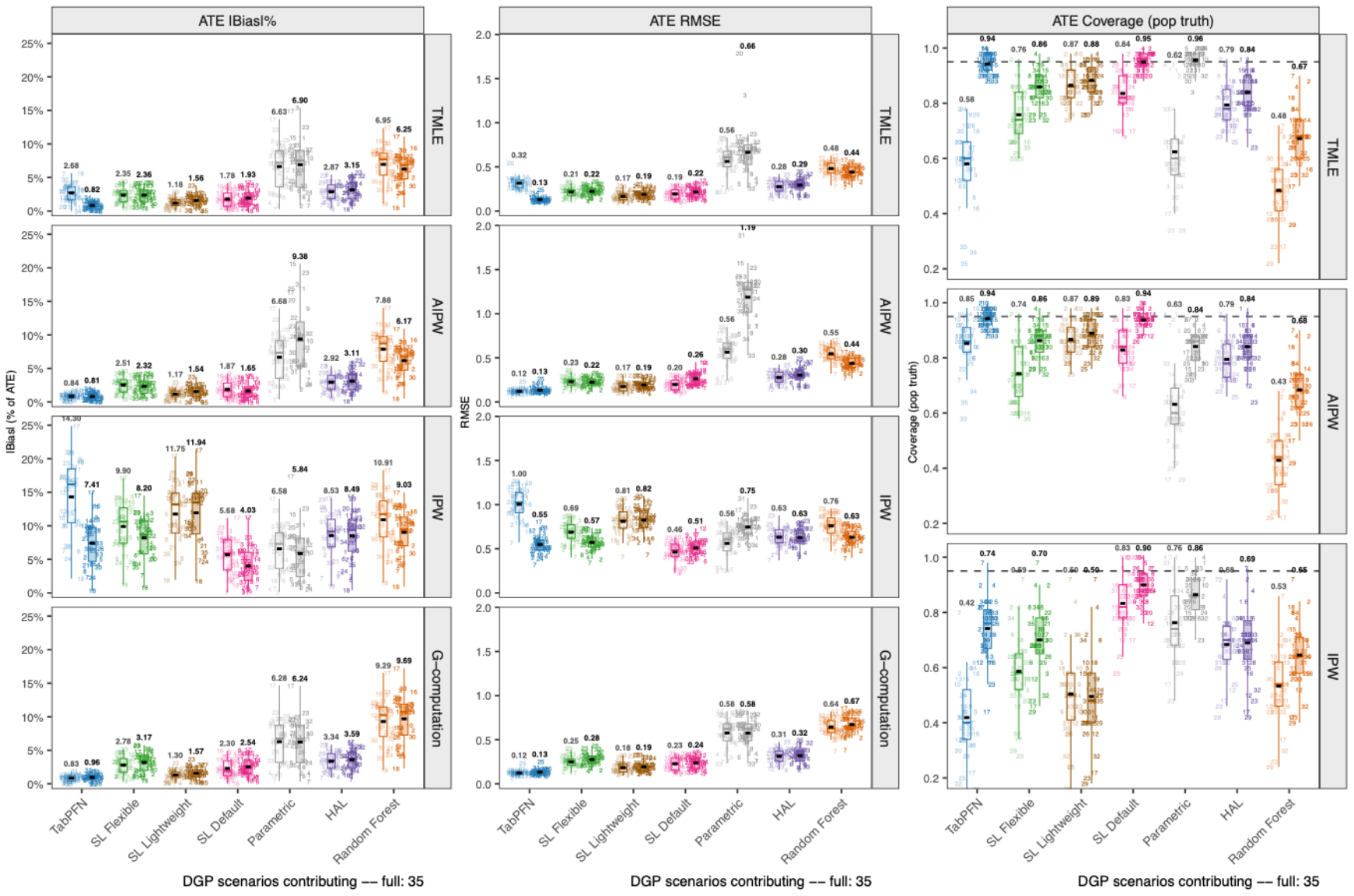

**Figure S3. ATE – Bias, RMSE, and coverage across all estimators and cross-fitting configurations from simulation 2.** Across 35 ATE-identified DGPs with full overlap. For each nuisance learner, the lighter points (left) are without cross-fitting and the darker points (right) are with cross-fitting. Black bars and bold labels indicate unweighted macro-averages across all contributing DGP settings. SEs and confidence intervals were not available for the G-computation estimator.

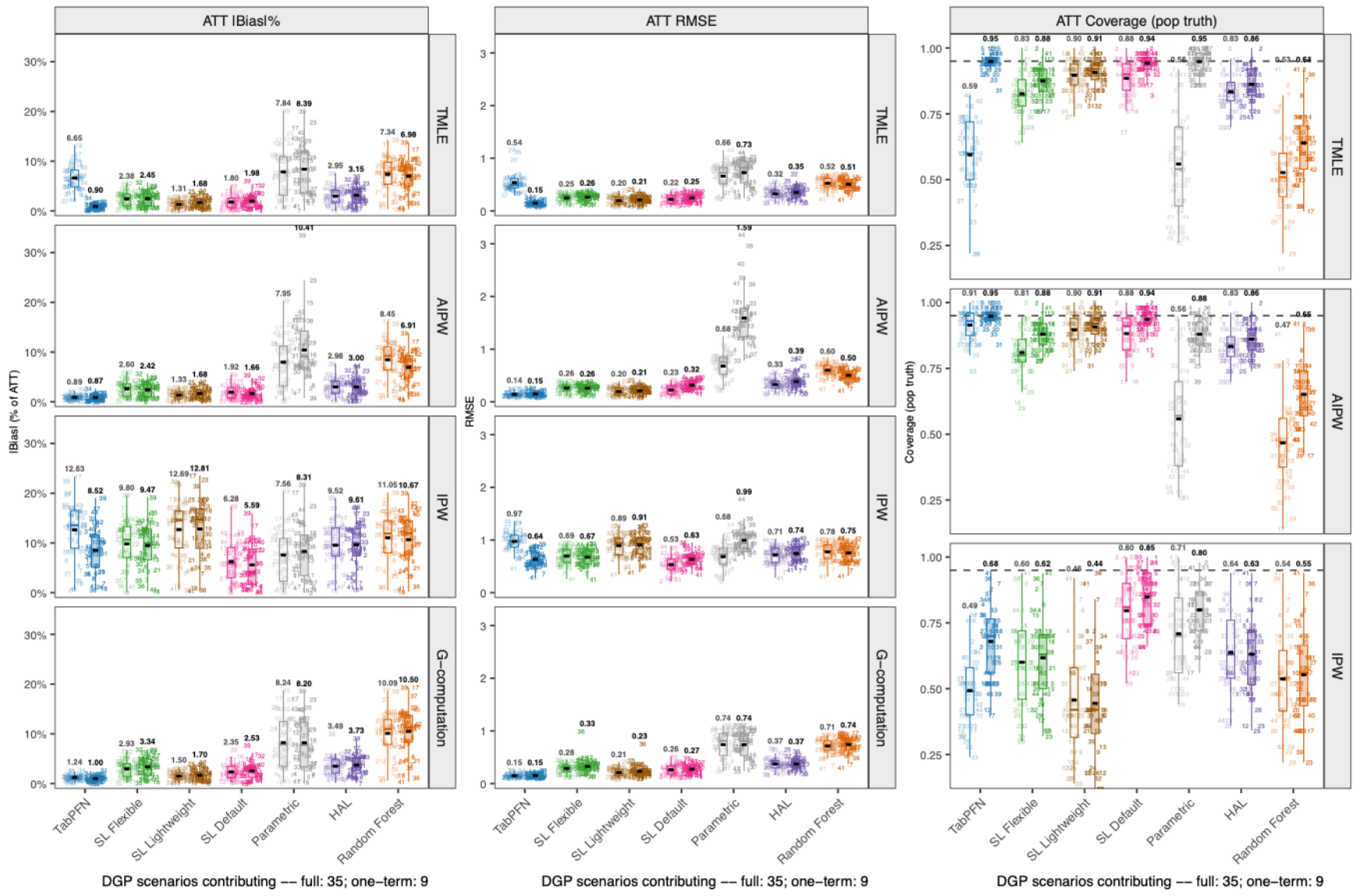

**Figure S4. ATT – Bias, RMSE, and coverage across all estimators and cross-fitting configurations from simulation 2.** Across 44 ATT-identified DGPs, including full and partial (one-term) overlap settings. For each nuisance learner, the lighter points (left) are without cross-fitting and the darker points (right) are with cross-fitting. Black bars and bold labels indicate unweighted macro-averages across all contributing DGP settings. SEs and confidence intervals were not available for the G-computation estimator.

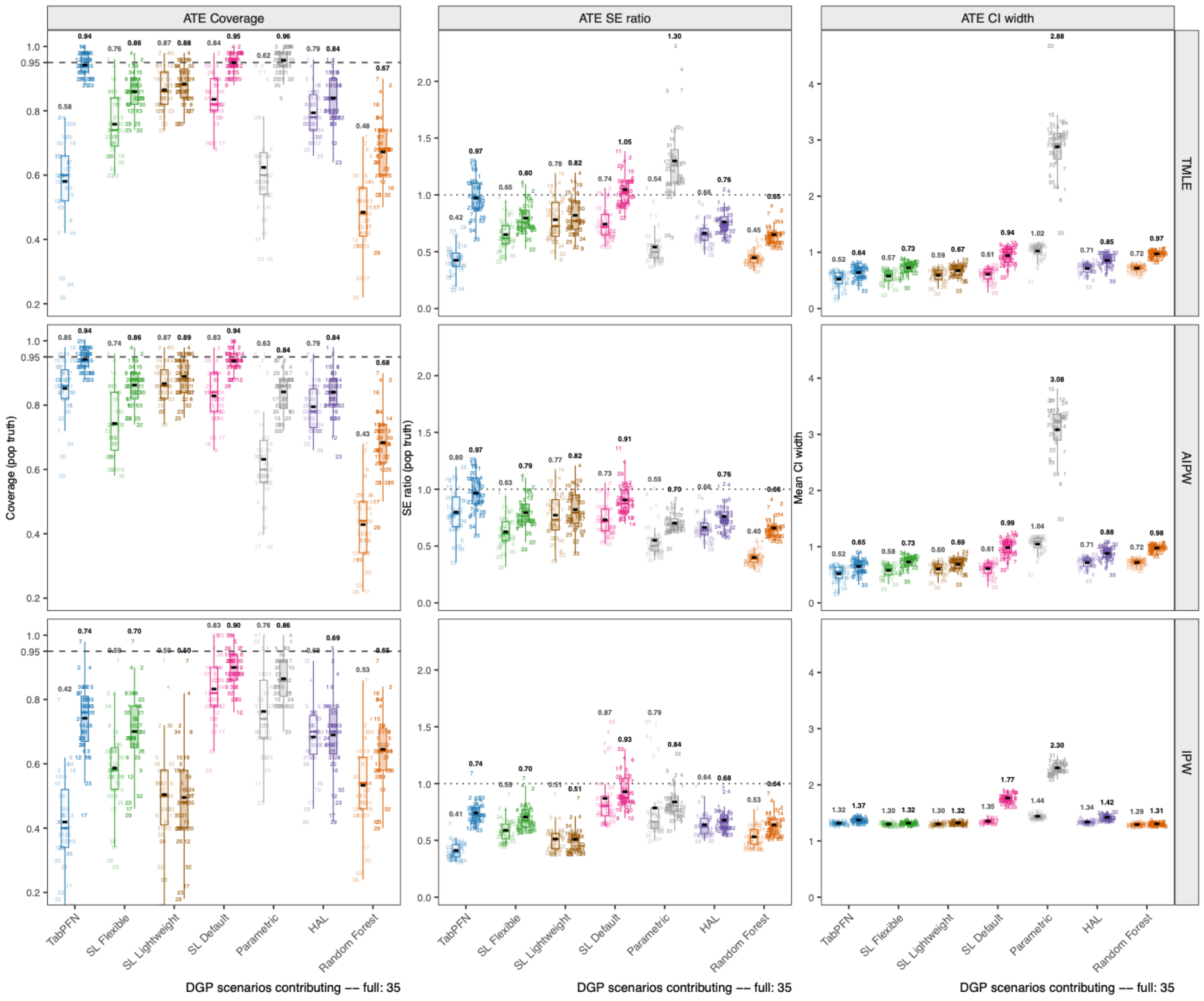

**Figure S5. ATE – Coverage and SE calibration for TMLE, AIPW, and IPW estimators from simulation 2.** Across 35 ATE-identified DGPs with full overlap. For each nuisance learner, the lighter points (left) are without cross-fitting and the darker points (right) are with cross-fitting. Black bars and bold labels indicate unweighted macro-averages across all contributing DGP settings. SEs and confidence intervals were not available for the G-computation estimator.

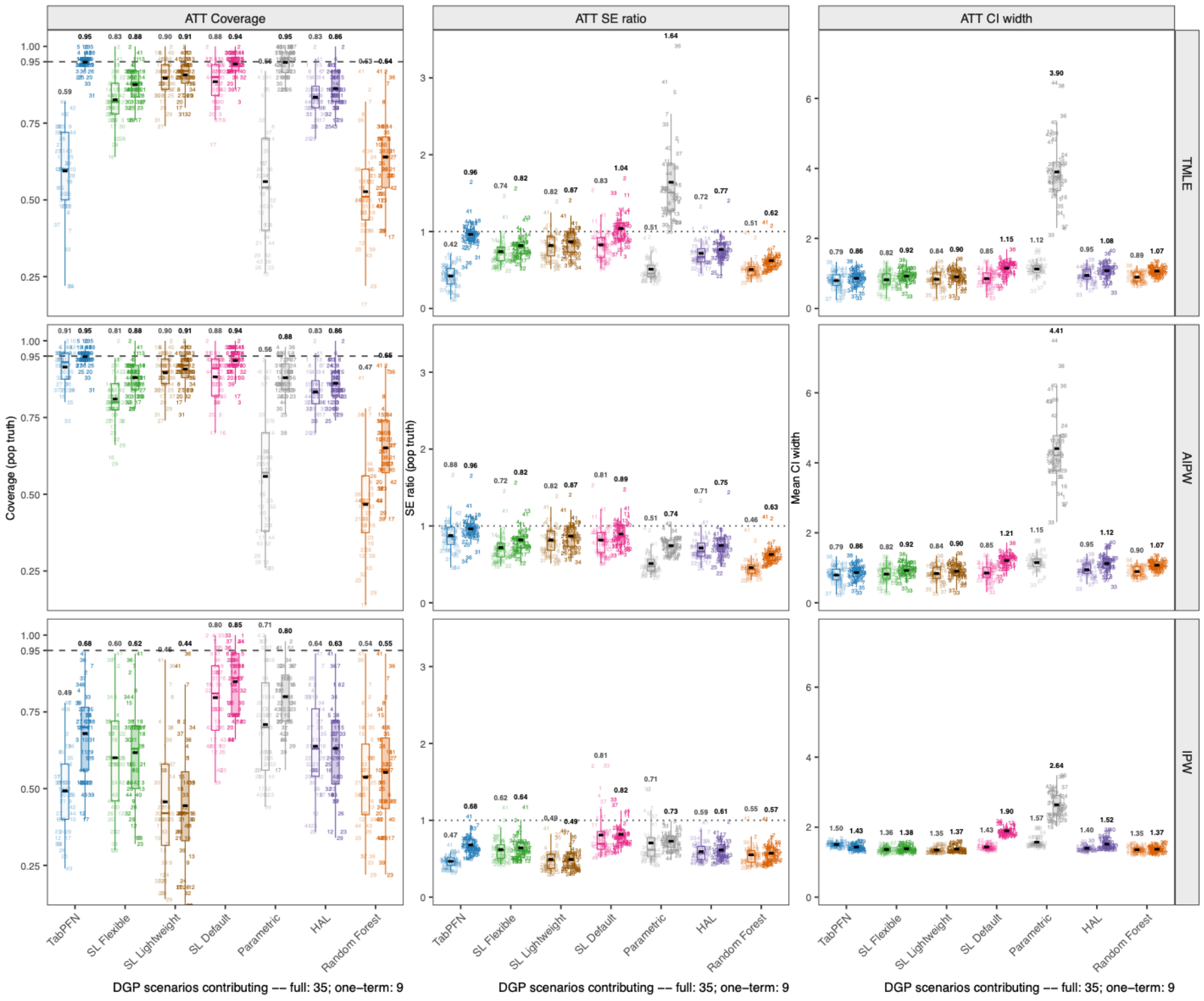

**Figure S6. ATT – Coverage and SE calibration for TMLE, AIPW, and IPW estimators from simulation 2.** Across 44 ATT-identified DGPs, including full and partial (one-term) overlap settings. For each nuisance learner, the lighter points (left) are without cross-fitting and the darker points (right) are with cross-fitting. Black bars and bold labels indicate unweighted macro-averages across all contributing DGP settings. SEs and confidence intervals were not available for the G-computation estimator.